\RequirePackage{fix-cm}
\documentclass[natbib]{svjour3}                     % onecolumn (standard format)
\smartqed  % flush right qed marks, e.g. at end of proof
\usepackage{graphicx}
\usepackage{epsf}
\usepackage{amssymb}
\usepackage{amsmath}
\usepackage{placeins}
\usepackage{color}
\usepackage{url}

 \journalname{SSRv}
\newcommand{\Msun}{~M_\odot}
\newcommand{\Rsun}{~R_\odot}
\newcommand{\msun}{M_\odot}

\newcommand{\kms}{\rm ~km~s^{-1}}

\begin{document}
\title{Superluminous supernovae}

\titlerunning{Superluminous supernovae}        % if too long for running head

\author{
Takashi~J.~Moriya\footnote{NAOJ Fellow} \and Elena~I.~Sorokina \and Roger~A.~Chevalier\\
%%(temporally in alphabetical order)
}

\authorrunning{T.J. Moriya, E.I. Sorokina, R.A. Chevalier} % if too long for running head

\institute{T. J. Moriya \at
              Division of Theoretical Astronomy,
              National Astronomical Observatory of Japan,
              National Institutes of Natural Sciences,
              2-21-1 Osawa, Mitaka, Tokyo 181-8588, Japan \\
              \email{takashi.moriya@nao.ac.jp}
           \and
           E. I. Sorokina \at
              Sternberg Astronomical Institute,
              M.V. Lomonosov Moscow State University,
              Universitetski pr. 13, 119234 Moscow, Russia \\
              \email{sorokina@sai.msu.su}
           \and
           R. A. Chevalier \at
              Department of Astronomy, University of Virginia,
              P.O. Box 400325, Charlottesville, VA 22904-4325, USA \\
              \email{rac5x@virginia.edu}
           }

\date{Received: 25 December 2017 / Accepted: 5 March 2018}

\maketitle

\begin{abstract}
Superluminous supernovae are a new class of supernovae that were recognized about a decade ago. Both observational and theoretical progress has been significant in the last decade. In this review, we first briefly summarize the observational properties of superluminous supernovae. We then introduce the three major suggested luminosity sources to explain the huge luminosities of superluminous supernovae, i.e., the nuclear decay of $^{56}$Ni, the interaction between supernova ejecta and dense circumstellar media, and the spin down of magnetars. We compare these models and discuss their strengths and weaknesses.
\end{abstract}

\section{Introduction}\label{sec:introduction}
Superluminous supernovae (SLSNe) are supernovae (SNe) that become more luminous than $\sim -21$~mag in optical. They are more than 1~mag more luminous than broad-line Type~Ic SNe, or the so-called ``hypernovae,'' which have kinetic energy of more than $\sim 10^{52}~\mathrm{erg}$ and are the most luminous among the classical core-collapse SNe. The first glimpse of their existence was in SN~1999as \citep{knop1999sn1999as}.  Other SLSNe, such as SCP06F6 \citep{barbary2009scp06f6}, were subsequently discovered. However, their nature was not clarified until SLSNe started to be discovered regularly with unbiased transient surveys such as that conducted by the ROTSE-IIIb telescope \citep{quimby2006rotseiiib}. Nowadays, SLSNe are discovered even at redshifts around 2 and beyond \citep{cooke2012highzslsn,pan2017slsnz1p861,smith2017z2des,moriya2018,curtin2018}, and they are becoming an important probe of, e.g., massive star formation in the high-redshift universe. They may be eventually applicable to cosmology like Type~Ia SNe \citep{quimby2013slsnrate,inserra2014cosmology,2018A&A...609A..83I,scovacricchi2016slsncosmology,wei2015slsncosmology}. They can also be a back light source to study interstellar media \citep{berger2012slsnism}.

This review mainly focuses on theoretical aspects of the energy sources of SLSNe. Three energy sources are mainly suggested to account for the huge luminosities of SLSNe, i.e., $^{56}$Ni decay (Section~\ref{sec:56ni}), interaction between ejecta and dense circumstellar media (CSM, Section~\ref{sec:interaction}), and magnetar spin-down (Section~\ref{sec:magnetars}). We start this review with a brief summary of the observational properties in Section~\ref{sec:obs} and then discuss the theoretical aspects of the power sources. We compare three models in Section~\ref{sec:comparison}. For a review of observational aspects of SLSNe, see, e.g., \citet{gal-yam2012slsnreview} and \citet{howell2017slsnreview}. Observational data on SLSNe are now available at \url{https://slsn.info/}, which is maintained by T.-W. Chen.

\section{Observational properties}\label{sec:obs}
Broadly speaking, SLSNe can be classified in two different spectral types; those with narrow hydrogen emission lines (Type~IIn, e.g., \citealt{smith2010sn2006gyspectra}) and those without \citep[e.g.,][]{quimby2011slsnic}. Most of SLSNe without the narrow hydrogen emission lines are classified as Type~Ic and we simply call them SLSNe~Ic. However, some SLSNe show broad hydrogen lines \citep[e.g.,][]{inserra2016slsnbroadh} and there exist some SLSNe that are initially Type~Ic but start to show hydrogen emissions about 1~year after the luminosity peak \citep{yan2015iptf13ehe,yan2017slsniclateh}. We introduce Type~IIn SLSNe in Section~\ref{sec:slsniin} and then the other SLSNe in Section~\ref{sec:slsnic}. We discuss the event rates of SLSNe in Section~\ref{sec:eventrate} and their environment in Section~\ref{sec:environment}.

\subsection{Type~IIn SLSNe}\label{sec:slsniin}
\begin{figure}
  \includegraphics[width=\textwidth]{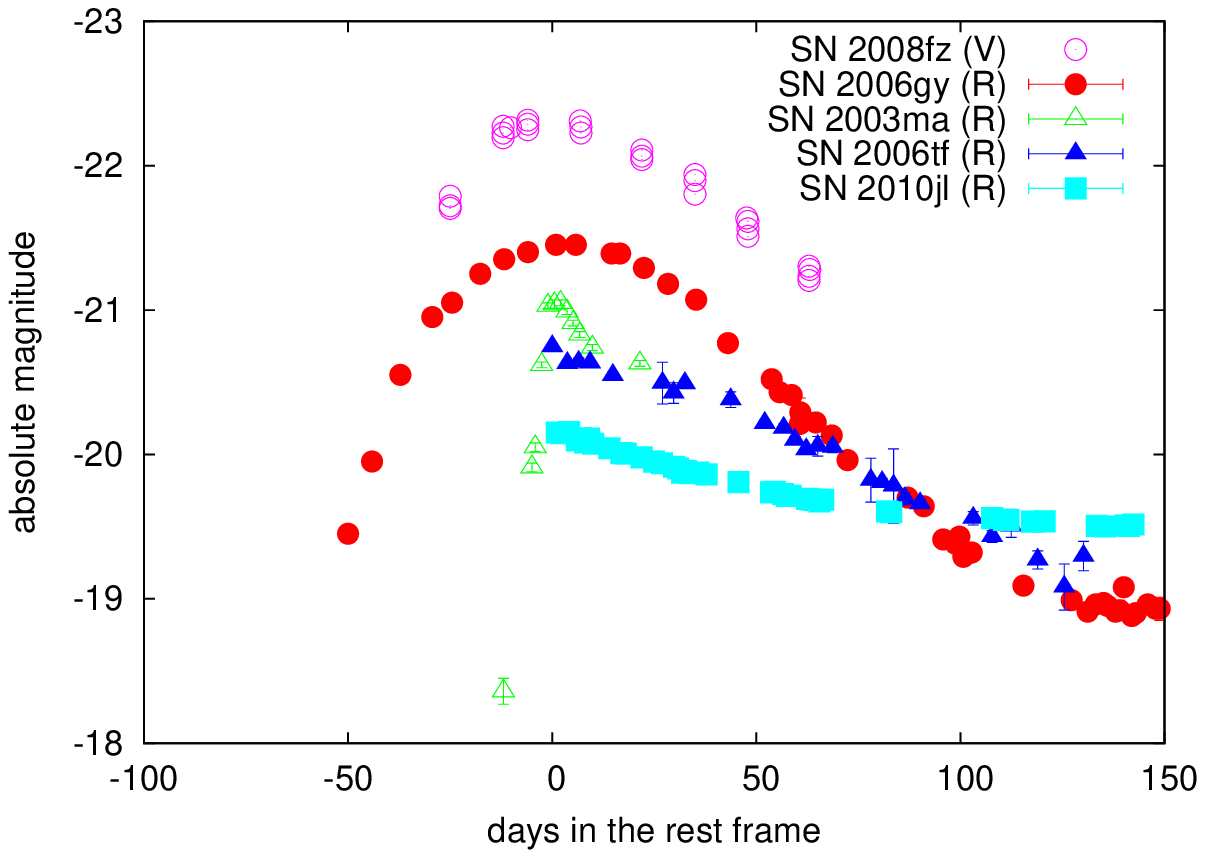} 
  \includegraphics[width=1\textwidth]{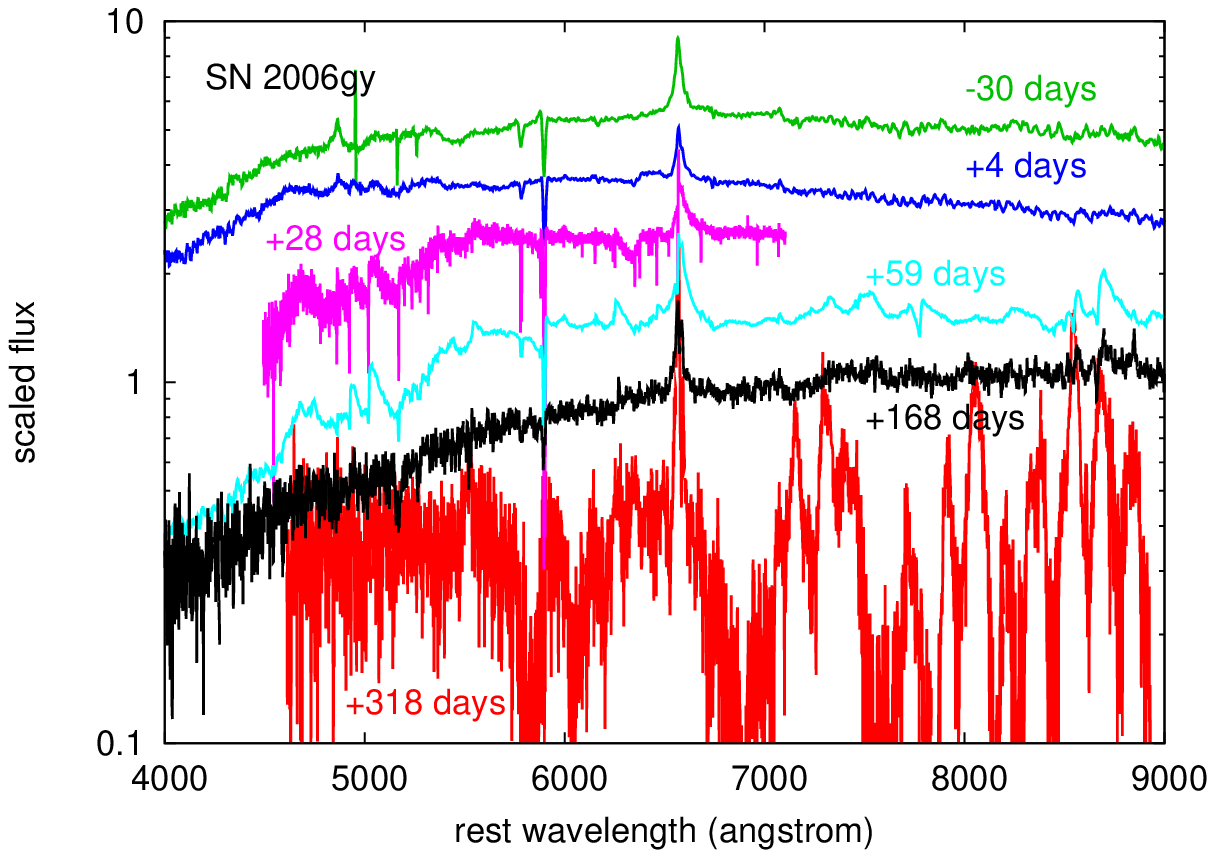}
\caption{
\textit{Top:}
LC diversities in Type~IIn SLSNe. The SN names and observed bands are indicated in the figure. The data sources are: \citet{drake2010sn2008fz} (SN~2008fz), \citet{smith2007sn2006gy} (SN~2006gy), \citet{rest2011sn2003ma} (SN~2003ma), \citet{smith2008sn2006tf} (SN~2006tf), and \citet{zhang2012AJsn2010jl} (SN~2010jl).
\textit{Bottom:}
Spectroscopic evolution of SLSN IIn 2006gy \citep[e.g.,][]{smith2010sn2006gyspectra,kawabata2009sn2006gy}. The dates are relative to the luminosity peak in the $R$ band in the rest frame.
}
\label{fig:typeiin}       
\end{figure}

SLSNe~IIn are characterized by narrow emission lines (Fig.~\ref{fig:typeiin}). Narrow hydrogen emission lines are particularly strong. The spectral characteristics of SLSNe~IIn are similar to those of less luminous SNe~IIn. Therefore, they are believed to be extreme cases of SNe~IIn where their luminosities are mainly powered by the interaction between ejecta and dense CSM (Section~\ref{sec:interaction}). SN~2006gy is the closest and the best observed SN in this class \citep[e.g.,][]{ofek2007sn2006gy,smith2007sn2006gy,smith2008sn2006gylate,smith2010sn2006gyspectra,kawabata2009sn2006gy,agnoletto2009sn2006gy,miller2010sn2006gyir}.

Although the  light curves (LCs) of SLSNe~IIn share the characteristics that they become extremely luminous, their LC shapes are diverse (Fig.~\ref{fig:typeiin}). SN~2008fz \citep{drake2010sn2008fz} and SN~2006gy \citep{smith2007sn2006gy} have broad round LCs lasting for more than 100~days while SN~2003ma \citep{rest2011sn2003ma} has a short round phase followed by a long phase of constant luminosity. The LCs of less luminous Type~IIn SNe like SN~2006tf \citep{smith2008sn2006tf} and SN~2010jl \citep{zhang2012AJsn2010jl,fransson2014sn2010jl} often decline with a power law. These diversities are likely to originate from the diversity in the CSM related to the diversity in the mass loss processes of the progenitors. 

It is not yet clear if SLSNe~IIn are just the most luminous end of a SN~IIn luminosity function or SLSNe~IIn make a separate population of their own \citep[e.g.,][]{richardson2014lumfunc}. There exist several SNe~IIn whose peak luminosity is between $\sim -20~\mathrm{mag}$ and $\sim -21~\mathrm{mag}$. These luminous populations can be observed more easily but there do not exist many SNe~IIn with which we can make a volume-limited sample because of their small event rate. The most luminous SLSN~IIn reported so far is CSS100217:102913+404220 reaching $-23$~mag in the optical at the LC peak (\citealt{drake2011css100217}, see also \citealt{kankare2017agntransients}). Although CSS100217:102913+404220 and other similar objects are suggested to originate from stellar explosions, they appear at the centers of galaxies with active galactic nuclei and their stellar explosion origin is doubted \citep[e.g.,][]{blanchard2017ps16dtm,moriya2017agn}. The most luminous published SLSN~IIn without a clear association with an active galactic nucleus is SN~2008fz \citep{drake2010sn2008fz}.

\subsection{Type~Ic SLSNe}\label{sec:slsnic}
\begin{figure}
  \includegraphics[width=\textwidth]{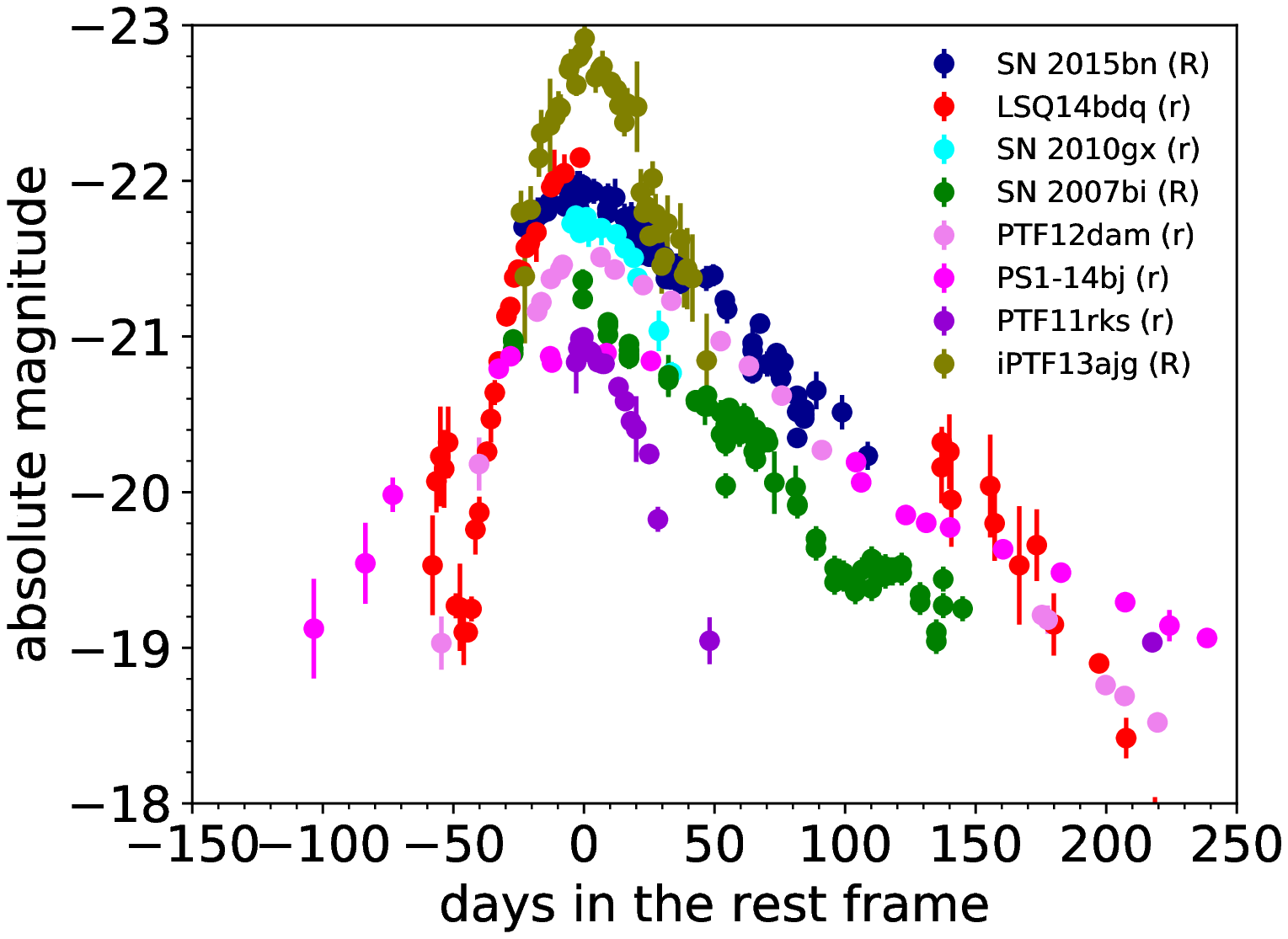} 
  \includegraphics[width=1\textwidth]{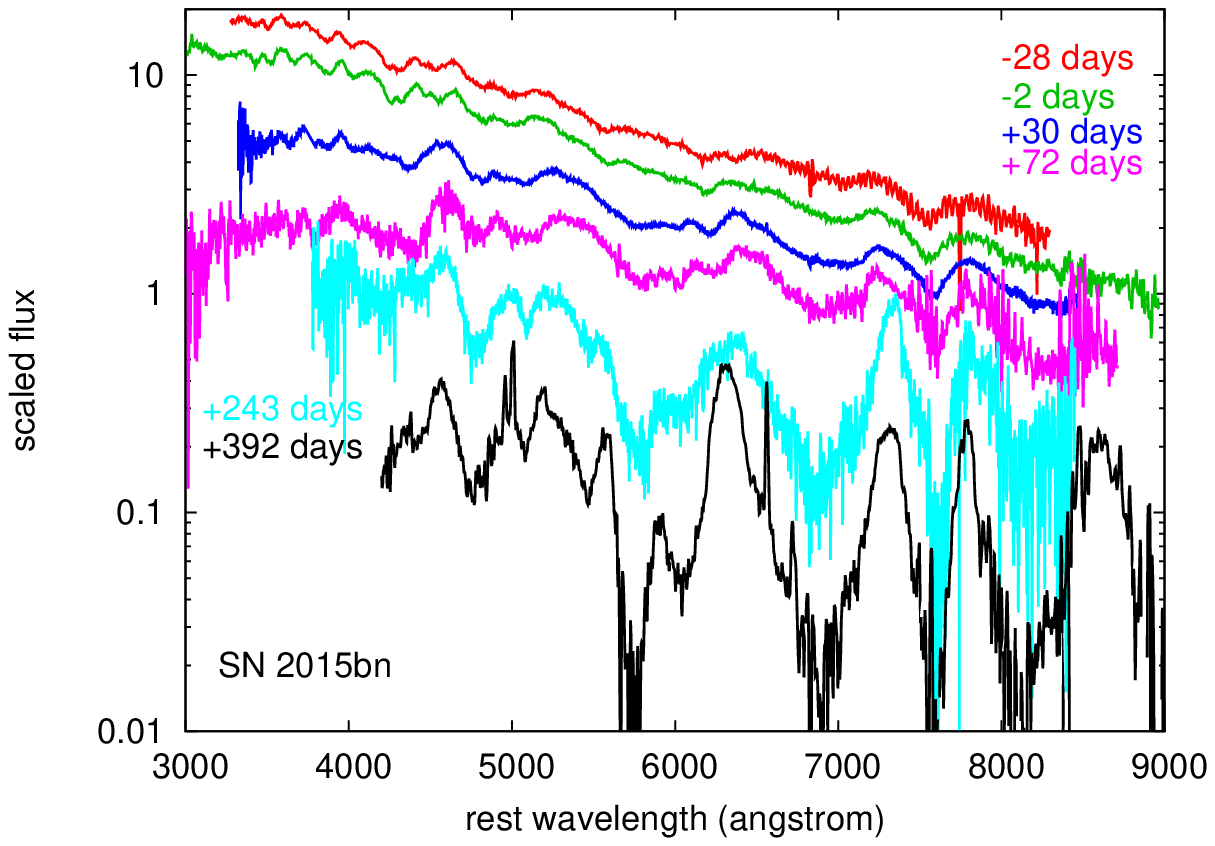}
\caption{
\textit{Top:}
LC diversities in Type~Ic SLSNe. The SN names and observed bands are indicated in the figure. The data sources are: \citet{nicholl2016sn2015bnearly} (SN~2015bn), \citet{nicholl2015lsq14bdq} (LSQ14bdq), \citet{pastorello2010sn2010gx} (SN~2010gx), \citet{gal-yam2009sn2007bi} (SN~2007bi), \citet{lunnan2016ps1-14bj} (PS1-14bj), \citet{inserra2013firstmagnetar} (PTF11rks), and \citet{vreeswijk2014iptf13ajg} (iPTF13ajg).
\textit{Bottom:}
Spectroscopic evolution of SLSN Ic 2015bn \citep{nicholl2016sn2015bnearly,nicholl2016sn2015bnnebular}. The dates are relative to the optical luminosity peak in the rest frame.
}
\label{fig:typeic}       
\end{figure}

SLSNe~Ic, which are often simply called SLSNe~I, are SLSNe without hydrogen spectral features around the time of the luminosity peak \citep[e.g.,][]{quimby2011slsnic}. Fig.~\ref{fig:typeic} shows LCs and spectra of SLSNe~Ic. The peak optical magnitudes of SLSNe~Ic are typically between $\sim -21~\mathrm{mag}$ and $-22~\mathrm{mag}$ \citep[e.g.,][]{nicholl2015slsndiversity,decia2017ptfslsnlc,lunnan2017slsnicps1}. The distribution of their peak bolometric magnitudes derived from optical LCs peaks at $-20.7~\mathrm{mag}$ with a small dispersion of 0.5~mag \citep{nicholl2015slsndiversity}. The luminosity function of Type~Ibc SNe peaks at around $-18$~mag with a dispersion of more than 0.5~mag \citep{drout2011snibc,taddia2015snibc,lyman2016snibc}. Therefore, there may exist a gap in the hydrogen-poor SN luminosity function between $\sim - 20$~mag and $\sim -21~\mathrm{mag}$, where a small number of Type~Ic BL supernovae and some gap transients exist \citep{arcavi2016gap,roy2016sn2012aa}. The rise times of SLSNe~Ic have a diversity, ranging from $\simeq 20~\mathrm{days}$ (PTF11rks, \citealt{inserra2013firstmagnetar}) to more than 125~days (PS1-14bj, \citealt{lunnan2016ps1-14bj}). However, their LC decline rates may have two distinct classes: rapidly-declining ones and slowly-declining ones \citep[e.g.,][]{nicholl2015slsndiversity}. The LC decline rates of slowly-declining SLSNe~Ic are often consistent with the nuclear decay rate of $^{56}\mathrm{Co}$ \citep[e.g.,][]{gal-yam2009sn2007bi,inserra2017slsnslowlcprop} and they are sometimes referred as SLSN~R, with ``R'' for radioactive decay \citep{gal-yam2012slsnreview}. However, whether they are powered by nuclear decay or not cannot be judged solely based on the LCs \citep[e.g.,][]{moriya2017nimimi}. The slowly-declining SLSNe~Ic may tend to be intrinsically more luminous than rapidly-declining SLSNe~Ic, which may enable us to use SLSNe~Ic as a distance indicator like Type~Ia SNe (\citealt{inserra2014cosmology}, but see also \citealt{quimby2013slsnrate}).

The LCs of SLSNe~Ic have a distinguishing feature at the very beginning. They often show a precursor before the main luminosity increase \citep[e.g.,][]{nicholl2015lsq14bdq,smith2016desslsnwithbump,vreeswijk2017slsnearlybump}. The possible existence of the precursor was first found in SN~2006oz \citep{leloudas2012sn2006oz} and it was systematically studied by \citet{nicholl2016slsnbump}. The precursor lasts for about $5-10$~days and its magnitude is around $-20$~mag. Its blackbody temperature is around 30,000~K \citep{nicholl2016slsnbump}. The precursor is followed by a fading lasting for about 5~days and the luminosity is about 1~mag fainter than the precursor luminosity in this period. Then, the LC starts to rise again to be a SLSN.

LC behaviors of slowly-declining SLSNe~Ic during the fading phase after the LC peak have been intensively studied by \citet{inserra2017slsnslowlcprop}. The LCs show fluctuations that may originate from the interaction with the CSM. The LC decline rates are consistent with that of the $^{56}$Co decay at first, but tend to decline faster after a few hundred days after the LC peak.

The early-time spectra of SLSNe~Ic at around the luminosity peak are characterized by a series of broad carbon and oxygen features \citep{quimby2011slsnic,quimby2018,chomiuk2011panslsn,howell2013snlsslsn,nicholl2014slsnfrompessto,nicholl2016sn2015bnearly,yan2017slsniuvspec,yan2017sn2017egmuv,liu2017slsn}. The blackbody temperature is around 15,000~K at these epochs, but non-thermal excitations play essential roles in forming the spectral features \citep[e.g.,][]{mazzali2016slsnicsp}. Photospheric velocities measured by using Fe~II $\lambda 5169$ are typically around $12,000~\mathrm{km~s^{-1}}$ and they do not evolve fast \citep[e.g.,][]{nicholl2015slsndiversity}. It is suggested that the existence of helium is required for the line formation \citep{mazzali2016slsnicsp} and a helium line may have been observed in SN~2012il \citep{inserra2013firstmagnetar}.  At later phases, the spectra often start to be similar to those of broad-line Type~Ic SNe \citep{pastorello2010sn2010gx,nicholl2016sn2015bnnebular}. In addition, there are several SLSNe~Ic that do not show hydrogen features near the LC peak but hydrogen emission lines start to appear about 1~year after the LC peak \citep{yan2015iptf13ehe,yan2017slsniclateh}. The late-phase hydrogen emissions have been suggested to originate from a detached hydrogen-rich circumstellar shell \citep{yan2015iptf13ehe,yan2017slsniclateh} or matter stripped from the progenitor's hydrogen-rich companion star \citep{moriya2015slsnhstri}.

There are a few SLSNe in which broad hydrogen features are observed \citep{gezari2009sn2008es,miller2009sn2008es,inserra2016slsnbroadh}. They usually do not show narrow features in spectra like SLSNe~IIn at early times. Therefore, their powering source may be related to SLSNe~Ic. However, their spectra start to show possible signatures of the interaction in the late phase and some luminosity contributions from the interaction may exist after around 1~year since the LC peak. The SLSN CSS121015:004244+132827 showed both broad and narrow hydrogen features from early times \citep{benetti2014css121015}.

Polarization of SLSNe~Ic was measured in three objects: LSQ14mo, SN~2015bn, and SN~2017egm. The first attempt was made for LSQ14mo. \citet{leloudas2015lsq14mopolari} measured photometric polarization in the $V$~band for 5 epochs and they did not detect clear polarization signatures. SN~2015bn appeared closer and both imaging and spectroscopic polarizations were measured for several epochs \citep{leloudas2017sn2015bnpolari,inserra2016sn2015bnpolari}. Aspherical shapes may be required to explain the polarization features. SN~2017egm is the closest SLSN~Ic and a spectropolarimetry observation was reported \citep{bose2017sn2017egm}. Polarization of about 0.5\%  without a strong dependence on wavelength is observed, indicating a modest departure from spherical symmetry. More observations are needed to obtain a systematic view of the asphericity in SLSNe~Ic. Polarization of one SLSN~II with broad hydrogen features is observed but a meaningful constraint on its shape could not be made \citep{inserra2016slsnbroadh}.

SLSNe~Ic are also actively observed in X-rays. SCP06F6 is the first SLSN~Ic with a possible detection of X-rays \citep{GaensickeXray06F6,levan2013slsnxray}. It was detected in X-rays ($0.1-2$~keV) by the \textit{XMM-Newton} satellite. The X-ray luminosity reached $\sim 10^{45}~\mathrm{erg~s^{-1}}$ in one epoch and then it was not detected any more. Another case is PTF12dam \citep{margutti2017slsnxray}. It was detected by \textit{Chandra} at $0.5-2~\mathrm{keV}$ in a similar epoch to SCP06F6 but its luminosity is $\sim 10^{40}~\mathrm{erg~s^{-1}}$ and is 5 orders of magnitude below that of SCP06F6. The X-ray luminosity at the location of PTF12dam is consistent with that expected from the underlying star-formation activities at the SN location and the X-rays may not originate from PTF12dam itself. X-ray observations in many other SLSNe~Ic have been performed but SCP06F6 and PTF12dam are the only cases with X-ray detections \citep{margutti2017slsnxray}. No radio emission is observed in SLSNe so far \citep{coppejans2017slsnradio}.

Gamma-ray bursts (GRBs) are often associated with Type~Ic-BL SNe. It is interesting to note that SLSNe~Ic have spectra similar to SNe~Ic-BL at late phases \citep[e.g.,][]{pastorello2010sn2010gx,nicholl2016sn2015bnnebular}. In addition, SN~2011kl, which was associated with an ultra-long GRB111209A, had a LC peak at $-20~\mathrm{mag}$ and it was much more luminous than SNe usually associated with GRBs \citep{greiner2015ulgrbslsn}. Although SN~2011kl was not as luminous as SLSNe and its spectroscopic features, which are overall featureless, do not appear to be similar to SLSNe~Ic, some GRBs may be related to SLSNe. \citet{yu2017flare} speculate that LC modulations in SN~2015bn may be related to flare activity of GRBs. Similarities in GRBs and SLSNe are also suggested in their host environment which will be discussed in a following section.

The most luminous SLSN~Ic reported so far is ASASSN-15lh, which reached about $-23.5$~mag  \citep{dong2016asasn15lh}. However, it appeared at the center of the host galaxy and its origin is questioned \citep{leloudas2016asasn15lh}. Its host galaxy is massive and metal-rich and, therefore, is peculiar %%%if it has a peculiar environment       RC
if it is actually a SLSN~Ic \citep[e.g.,][]{leloudas2016asasn15lh}.
Except for ASASSN-15lh, the most luminous SLSNe~Ic reach $ -22.5$ $\sim$ $-23$~mag in optical (e.g., PTF13ajg, \citealt{vreeswijk2014iptf13ajg}).

\subsection{Event rates}\label{sec:eventrate}
\citet{quimby2013slsnrate} calculate the rates of SLSNe at $z\sim 0.2$. They find a total SLSN rate of $199^{+137}_{-86}~\mathrm{Gpc^{-3}~yr^{-1}} h_{71}^3$, where $h_{71}$ is the Hubble constant scaled with $71~\mathrm{km~s^{-1}~Mpc^{-1}}$. They estimate a SLSN~II rate of $151^{+151}_{-82}~\mathrm{Gpc^{-3}~yr^{-1}} h_{71}^3$. Their SLSN~II sample consists of SLSNe~IIn, with the exception of SN~2008es. The rate of SLSNe~Ic is estimated to be  $32^{+77}_{-26}~\mathrm{Gpc^{-3}~yr^{-1}} h_{71}^3$. Because the core-collapse SN rate at $z\sim 0.2$ is $\sim 10^5~\mathrm{Gpc^{-3}~yr^{-1}}$ \citep[][]{madau2014rate}, the fraction of SLSNe to core-collapse SNe are estimated to be $\sim 10^{-3}$. The total SLSN rate is similar to the long GRB rate \citep[e.g.,][]{soderberg2006grb060218} and the SLSN~Ic  rate may be about 10\% of the long GRB rate.

There are two works deriving the SLSN~Ic rate at $z\sim 1$. The first work is by \citet{mccrum2015slsnpanst} who use a SN sample from the Pan-STARRS survey to derive a relative fraction of SLSNe~Ic to core-collapse SNe. They found only $\sim 10^{-5}$ of core-collapse SNe are SLSNe~Ic at $z\sim 1$, which is roughly a factor of 10 smaller than the rate at $z\sim 0.2$. The subsequent work by \citet{prajs2017rate} measured the SLSN~Ic rate at $z\sim 1$ based on the SLSN~Ic samples from the Supernova Legacy Survey. The survey has two spectroscopically confirmed SLSNe~Ic at $z=1.50$ and 1.59 and one additional SLSN~Ic candidate at $z=0.76$ which is found by fitting the LCs found in the survey. The three SLSNe~Ic in the survey give a SLSN~Ic rate of $91^{+76}_{-36}~\mathrm{Gpc^{-3}~yr^{-1}} h_{70}^3$ at $z\sim 1$. This rate is equivalent to $\sim 10^{-4}$ of the core-collapse rate at $z\sim 1$ and an order of magnitude higher than that estimated by \citet{mccrum2015slsnpanst}.

\citet{cooke2012highzslsn} reported two SLSNe at $z=2.05$ and $z=3.90$. Although their spectral types are not clear, the detection of two SLSNe in their survey very roughly corresponds to the SLSN rate of $\sim 400~\mathrm{Gpc^{-3}~yr^{-1}}h_{70}^3$ at $z=2-4$. The increase in the SLSN rate is still consistent with that expected from the increase in the cosmic star-formation rate and it is still not clear if the intrinsic SLSN frequency increases at high redshifts (see also a recent estimate by \citealt{moriya2018}).

\subsection{Environments}\label{sec:environment}
The host environments of SLSNe have been studied extensively. SLSNe~IIn are known to come from a broad range of metallicities \citep{neill2011slsnenv,perley2016slsnhost,schulze2016slsnenv,angus2016slsnhosthst}, while SLSNe~Ic only come from metal-poor environments \citep[e.g.,][]{schulze2016slsnenv,chen2016slsnhostmag,chen2015ptf12damhost,chen2013sn2010gx,leloudas2015slsnhost,perley2016slsnhost,vreeswijk2014iptf13ajg,lunnan2016ps1-14bj}. There is likely a suppression of SLSNe~Ic above around the half-solar metallicity environment \citep{perley2016slsnhost,chen2016slsnhostmag,schulze2016slsnenv}. SLSNe~Ic also tend to come from star-bursting galaxies with high specific star-formation rates \citep[e.g.,][]{schulze2016slsnenv}. This tendency can be explained if their progenitors are very massive stars \citep[e.g.,][]{thoene2015ptf12damhost,leloudas2015slsnhost}, but this interpretation is also questioned \citep[e.g.,][]{perley2016slsnhost}. SLSNe~Ic are often found in interacting galaxies where the star formation can be triggered by the interaction \citep[e.g.,][]{chen2016lsq14mo,cikota2017slsnhost}. SLSNe~Ic may prefer metal-poorer environments than long GRBs \citep[e.g.,][]{schulze2016slsnenv}. However, the differences between SLSNe~Ic and long GRB host environments are not yet clear \citep[e.g.,][]{lunnan2014slsngrbhosts,lunnan2015slsnhosthst,angus2016slsnhosthst,japelj2016slsngrbhosts}.

A SLSN~Ic SN~2017egm has appeared in a nearby spiral galaxy that has a high metallicity \citep{bose2017sn2017egm,nicholl2017sn2017egm} but the SN location itself may have low metallicity (\citealt{izzo2017sn2017egmenv}, but see also \citealt{chen2017sn2017egmhost}). %Further observations at the SN location are required to confirm the metallicity of the SN location.

An interesting relation between the host metallicities and the spin periods of magnetars required to fit the LCs of SLSNe~Ic has been suggested by \citet{chen2016slsnhostmag}. This relation may simply come from the relation between the host metallicities and total radiated energy in SLSNe~Ic, because the spin periods in the magnetar model represent the total available energy to power SLSNe (Section~\ref{sec:magnetars}). The suggested relation is still based on a small number of SLSNe~Ic and it needs to be studied more. Recent follow-up studies based on larger SLSN samples infer that the relation may not be as strong as originally suggested \citep{nicholl2017,decia2017ptfslsnlc}.

% 56Ni model
\section{$^{56}$Ni-powered models}\label{sec:56ni}
The standard source of early luminosity in SNe is the nuclear decay of $^{56}$Ni. The simplest approach to explain the huge luminosities of SLSNe is to increase the amount of $^{56}$Ni.
$^{56}$Ni decays to $^{56}$Co with a decay time of $8.76 \pm 0.01$~days \citep{dacruz1992} and then $^{56}$Co decays to $^{56}$Fe with a decay time of $111.42 \pm 0.04$~days \citep{funck1992}.
The total available energy from the nuclear decay is
\begin{equation}
\left[6.48\exp\left(-\frac{t}{8.76~\mathrm{days}}\right)
+1.44\exp\left(-\frac{t}{111.42~\mathrm{days}}\right)\right]\frac{M_\mathrm{^{56}Ni}}{M_\odot}10^{43}\mathrm{erg~s^{-1}},
\end{equation}
where $M_\mathrm{^{56}Ni}$ is the total mass of $\mathrm{^{56}Ni}$ available at the beginning (\citealt{nadyozhin1994}, but updated based on \url{http://www.nndc.bnl.gov/chart}). However, all the available energy from the decay is not necessarily absorbed by the SN ejecta to heat them up. All the decay energy of $\mathrm{^{56}Ni}$ and about 97\% of the decay energy of $\mathrm{^{56}Co}$ are released in the form of $\gamma$-rays. The released $\gamma$-rays travel in the ejecta and some fraction is absorbed by the ejecta and heats them up. The effective $\gamma$-ray opacity is $0.027~\mathrm{cm^2~g^{-1}}$ \citep{axelrod1980}. The remaining 3\% of the $\mathrm{^{56}Co}$ decay energy is released as positrons that can be all absorbed \textit{in situ} by the ejecta to heat them up. During the early phases when the ejecta are dense, most $\gamma$-rays can be absorbed. However, in the nebular phases when the ejecta are optically thin, a proper treatment of the $\gamma$-ray transport is needed to estimate the expected luminosity. Because heating by the $\mathrm{^{56}Ni}$ decay is a standard way to power SNe, we refer to a standard text such as \citet{arnett1996} for further details of the SN properties powered by $\mathrm{^{56}Ni}$ decay.

The required $^{56}$Ni masses to explain the peak luminosity of SLSNe are $5-20~M_\odot$ \citep[e.g.,][]{nicholl2015slsndiversity}.
There are several suggested ways to make such a large amount of $^{56}$Ni. One is the very energetic explosions of massive stars. \citet{umeda2008nomoto} investigated the explosive nucleosynthesis of very massive stars up to the zero-age main-sequence (ZAMS) mass of 100~$M_\odot$ with $Z_\odot/200$. They show that more than $5~M_\odot$ of $^{56}$Ni can be produced in massive cores if the explosion energy exceeds $10^{52}~\mathrm{erg}$. To produce more than $10~M_\odot$ of $^{56}$Ni, which is often required to reproduce the SLSN LCs, an unrealistically high explosion energy of $10^{53}~\mathrm{erg}$ is required. Relatively faint SLSNe requiring $\sim 5~M_\odot$ of $^{56}$Ni, like SN~2007bi, may be explained by the energetic core-collapse explosions of very massive stars \citep{young2010sn2007bi,moriya2010sn2007bi}.

A promising mechanism to synthesize more than $10~M_\odot$ of $^{56}$Ni in massive stars is through pair-instability \citep[e.g.,][]{rakavy1967pisn,barkat1967pisn}. If a star has a core that is massive enough, photons in the stellar core can be efficiently converted to electron and positron pairs and the center of the star can suffer a strong reduction in radiative pressure supporting the massive core. The reduction of the pressure support triggers the contraction of the massive core by making the adiabatic index below $4/3$ and  leads to explosive carbon and oxygen burning. If the energy released by the explosive nuclear burning is enough to unbind the whole star, it can explode without leaving any compact remnant. This kind of explosion is called pair-instability SN (PISN). The amount of $^{56}$Ni produced in PISNe strongly depends on the core mass of the progenitor and it ranges from almost zero to $\sim 50~M_\odot$ \citep[e.g.,][]{heger2002pisn,takahashi2016pisnnuc}. PISNe can occur if helium cores of massive stars are between $\sim 65~M_\odot$ and $\sim 135~M_\odot$ which corresponds to  ZAMS masses between $\sim 150~M_\odot$ and $\sim 250~M_\odot$ at zero metallicity \citep[e.g.,][]{heger2002pisn}. The ZAMS mass can be as low as $65~M_\odot$ if massive stars rotate rapidly \citep[e.g.,][]{chatzopoulos2012rotpisnprog,yusof2013massivestars}. The rapid rotation also helps PISN progenitors to remove their hydrogen through quasi-chemically homogeneous evolution \citep[e.g.,][]{yoon2012popiii}. It is interesting to note that most PISN candidates are hydrogen-free SLSNe (but see \citealt{2017NatAs...1..713T}), although hydrogen-rich PISNe can also be very luminous \citep[e.g.,][]{kasen2011pisn}. The surface instability of red supergiant PISN progenitors can also help reduce the amount of hydrogen in their explosions \citep{moriya2015langer}. PISNe are considered to exist in environments with metallicities less than $\sim Z_\odot/3$ \citep[e.g.,][]{langer2007pisn}, but they may exist even in a solar-metallicity environment if magnetic fields of massive stars can suppress their mass loss \citep{georgy2017magpisn}.

Predicted LCs of PISNe \citep[e.g.,][]{2005ApJ...633.1031S,kasen2011pisn,dessart2013pisn,whalen2014localpisn,kozyreva2014lowzpisn,chatzopoulos2015rotpisnlcsp} are roughly consistent with those of slowly-declining SLSNe~I whose late-phase decline rates are consistent with the $^{56}$Co nuclear decay rate. PISNe tend to have large rise times compared to SLSNe~I but some PISN models are consistent with the relatively short rise times of SLSNe~Ic \citep[e.g.,][]{kozyreva2017fastpisn}. The internal mixing of PISNe can help reduce their rise times as well \citep[e.g.,][]{kozyreva2015mixingpisn} but multidimensional simulations of PISN explosions generally do not find strong mixing in them \citep{joggerst2011pisnmixing,chen2014pisnmixing,gilmer2017pisnmixing}. Crucial issues of PISN models for SLSNe~Ic are in their spectroscopic properties. Although PISNe produce a huge amount of $^{56}$Ni, it is not sufficient to heat the massive core as much as observed \citep[e.g.,][]{dessart2012magni}. The production of a large amount of $^{56}$Ni also leads to the existence of a large amount of Fe-group elements in the ejecta. The absorptions and emissions by the Fe-group elements results in much redder spectra than those observed in SLSN~Ic \citep[e.g.,][]{dessart2012magni,jerkstrand2016pisnlate}.

Finally, we summarize pros and cons of the $^{56}$Ni-powered models.\\
Pros:
\begin{itemize}
 \item $^{56}$Ni power can naturally explain the decay rate of slowly declining SLSNe~Ic.
\end{itemize}
Cons:
\begin{itemize}
 \item $^{56}$Ni power cannot explain the rapidly declining SLSNe~Ic. A different model must be invoked for the rapidly declining SLSNe~Ic.
 \item Spectra of SLSNe~Ic do not match those predicted for $^{56}$Ni-powered SNe.
\end{itemize}

\section{Interaction-powered models}\label{sec:interaction}
Interaction of the gas ejected during a SN explosion with an ambient medium is a process that happens quite often at different stages of SN evolution.
Interaction leads to  shock wave formation, and shocks  very efficiently transform the  kinetic energy to  thermal energy, 
which can be then radiated 
at different wavelengths \citep{chevalier2003}.
The stage of SN evolution at which significant interaction  starts depends on the density of the CSM.
If the ejecta expand into low density material, the interaction becomes important at the SN remnant stage, a few tens, hundreds, or even thousands of years after the SN explosion.
Up to this age, the ejecta are cold and faint, but the shock interaction with the surrounding material heats it up again
\citep{Reynolds2016}.
%and observers can see very beautiful bright filaments (see the Chapter on SN remnants in this book). 
%Examples of the interaction with denser CSM can be found in the Chapter on Interaction. 
In the case of a dense CSM, the interaction occurs already on the SN stage 
\citep[e.g., for SNe of Type IIn, see][]{Chandra2018IIn}.
The SN becomes more luminous than a normal one exploded in lower density surroundings.

In the case of SLSNe,
the radiation energy emitted during the first few months after the explosion is of the order of $10^{50}$--$10^{51}$~ergs
\citep{quimby2011slsnic}.
The kinetic energy of a standard SN explosion is also $\sim 10^{51}$~ergs.
Naked SNe preserve almost all their kinetic energy until the remnant stage, 
when a sufficient amount of ambient mass (comparable to the ejecta mass) is swept up, 
and the interaction becomes important.
If the ambient mass is concentrated close to the exploding star,
the interaction starts from the very beginning and there is a good chance to transform the kinetic energy of the ejecta to radiation during the first months,
exactly as it is needed to produce a SLSN event.

%\begin{figure}
%\centering
%  \includegraphics[width=.5\textwidth]{CSMinteractionSketchChugai97.png} 
%\caption{{\bf Do we need this plot or some other illustration to clarify what the interaction model is?} 
%This sketch, from \citet{chugai1997}, roughly illustrates the model of the SN ejecta interacting with a clumpy CSM.
%}
%\label{fig:interactionSketch}       
%\end{figure}

In this Section we first make crude estimates of some parameters of the explosion  in order to understand what kind of structure can produce such a powerful event as a SLSN in the CSM interaction scenario (Sect.~\ref{subsec:InteractionParams}). 
Then we briefly discuss how the shock wave propagates through the dense extended CSM (Sect.~\ref{subsec:InteractionHydro}).
After that we describe analytical and numerical models for the SLSN light curves that originate from ejecta--CSM interaction and discuss the progenitors that are able to provide the required CSM structure (Sect.~\ref{subsec:InteractionLCs}).
Finally, we list pros and cons of the CSM interaction scenario for SLSNe (Sect.~\ref{subsec:InteractionProsCons}).

\subsection{Estimates of model parameters} \label{subsec:InteractionParams}
{\it Mass ratio $m_{\rm ej}/m_{\rm CSM}$ needed for SLSNe.} In the CSM interaction scenario for SLSNe, the expanding ejecta collide with the CS envelope that can be static or expanding. 
Let us consider for simplicity an inelastic collision of two blobs of matter.
The first one, which represents the ejecta, has mass $m_1$ and momentum $\mathbf{p_1}$.
Its kinetic energy is
\begin{equation}
   E_{\rm init} = \frac{ \mathbf{p_1}^2}{ 2 m_1} .
\end{equation}
It collides with another blob with mass $m_2$ and, for now, we suppose its momentum to be zero.
The final energy of two merged blobs in a fully inelastic collision is
\begin{equation}
   E_{\rm fin} = \frac{ \mathbf{p_1}^2}{2 (m_1+m_2)} .
\end{equation}
The momentum is conserved, but the energy $E_{\rm init}-E_{\rm fin}$ is  radiated away, since $E_{\rm fin} < E_{\rm init}$.
If $m_2 \ll m_1$ then only a tiny fraction of $E_{\rm init}$ is radiated, but if  $m_2 \gg m_1$ then
$E_{\rm fin} \ll E_{\rm init}$ and almost all initial kinetic energy of the system is radiated away.

\begin{figure}
\centering
  \includegraphics[width=.7\textwidth]{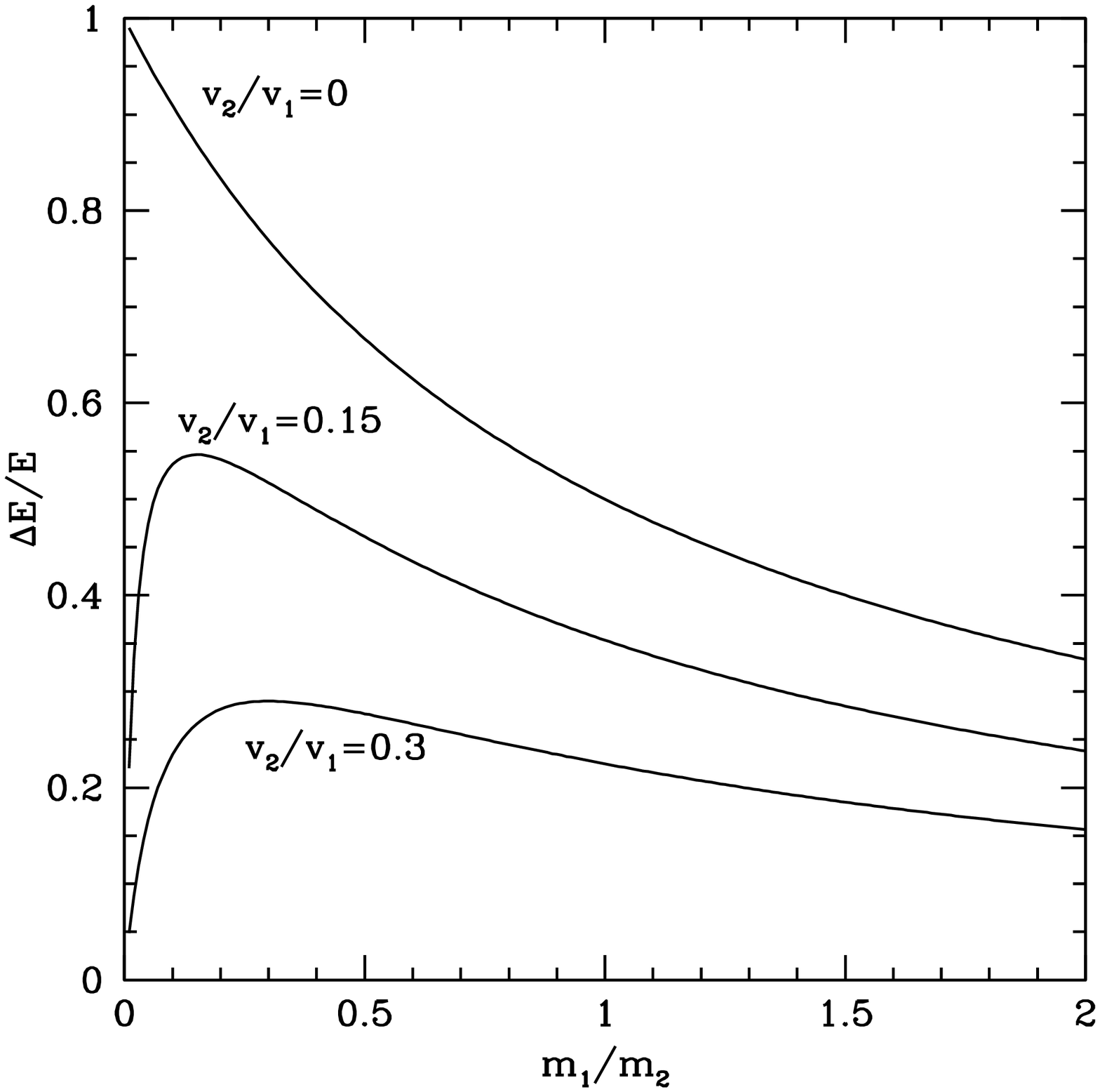} 
\caption{ Energy losses relative to the initial kinetic energy of the system of two inelastically colliding blobs.
}
\label{fig:Elosses}       
\end{figure}

The situation is a bit more complicated when the second blob, which represents the CSM envelope, is not static at the time of the explosion, 
but expands with a momentum $\mathbf{p_2}$.
This is most probable for the case of SLSNe I, because velocities higher than 10,000~km/s are observed for several months starting from the maximum
\citep{liu2017slsn}.
The energy before the interaction in this case is the sum of ejecta and CSM energies. 
In the prescription of two blobs, that is the sum of the blobs' energies:
\begin{equation}
   E_{\rm init} = \frac{ \mathbf{p_1}^2}{2 m_1} + \frac{ \mathbf{p_2}^2 }{2 m_2}.
\end{equation}
After the inelastic interaction the kinetic energy of the system becomes 
\begin{equation}
   E_{\rm fin} = \frac{ (\mathbf{p_1} + \mathbf{p_2})^2} { 2 (m_1+m_2)} .
\end{equation}
So an amount of energy that can be thermalized during the interaction is 
\begin{equation}
   \Delta E = E_{\rm init}-E_{\rm fin} = \frac { (\mathbf{p_1}m_2 - \mathbf{p_2}m_1)^2} { 2 m_1 m_2 (m_1+m_2)}
            = \frac{m_1 m_2} { 2 (m_1+m_2)} (\mathbf{v_1} - \mathbf{v_2})^2.
\end{equation}
Here $\mathbf{v_1}$ and $\mathbf{v_2}$ are velocities of the colliding blobs that represent, respectively, the ejecta and the CSM envelope.
Again, the important question is how much of the total kinetic energy of the system can be lost to radiation,
because for the superluminous event this energy transformation must be very effective.
If we define $\mu = m_1/m_2$ and $\beta = v_2/v_1$ then we get
\begin{equation}
   \frac{\Delta E} { E_{\rm init}} = \frac{\mu (1-\beta)^2} {(\mu+1)(\mu +\beta^2).}
\end{equation}
It is clear that $v_1>v_2$ ($\beta<1$). Otherwise, the collision will never happen.
This means that the larger the difference between $v_1$ and $v_2$, the larger the energy that can be radiated.
For a few fixed values of $\beta$, the dependence of $\Delta E/E_{\rm init}$ on the mass ratio $\mu$ is shown in  Fig.~\ref{fig:Elosses}.
When the CS envelope is expanding ($v_2>0$) the maximum losses happen not at $m_1\sim 0$, but at some finite mass.
Still, $m_1$ has to be noticeably lower than $m_2$ for the effective energy loss to happen.
This means that the ejecta must be less massive than the CS envelope for a SLSN event.

{\it Typical radius.} A crude estimate yields the photospheric radius $R_{\rm ph}$, and, therefore, the typical size of the dense circumstellar envelope
that can provide the observed luminosity $L$ of the SLSNe.
For simplicity, assume that the SLSN emits like a blackbody with a temperature $T_{\rm bb}$.
Then the photospheric radius can be obtained from 
$L = 4\pi R_{\rm ph}^2 \sigma T_{\rm bb}^4$.
The typical temperature measured for SLSNe is $T_{\rm bb}\simeq 10^4$~K.
The order of magnitude estimate of the luminosity is $L\simeq 10^{44}$~erg/s.
Hence, the radius is roughly $R_{\rm ph}=4\times 10^{15}$~cm, 
which is about an order of magnitude larger than typical photospheric radii of standard SNe near maximum light.

\citet{ginzburg2012csmlcmodel} in their simple numerical hydrodynamic diffusion model got a similar value for the CSM radius
(of order $10^{15}$~cm), 
and also stated that the masses of the ejecta and the CSM involved into the interaction must be comparable for the effective transformation of energy into radiation. 
As we see from the Fig.~\ref{fig:Elosses}, a smaller ejecta-to-CSM mass ratio is even better.

\begin{figure}
\centering
\includegraphics[width=0.74\linewidth]{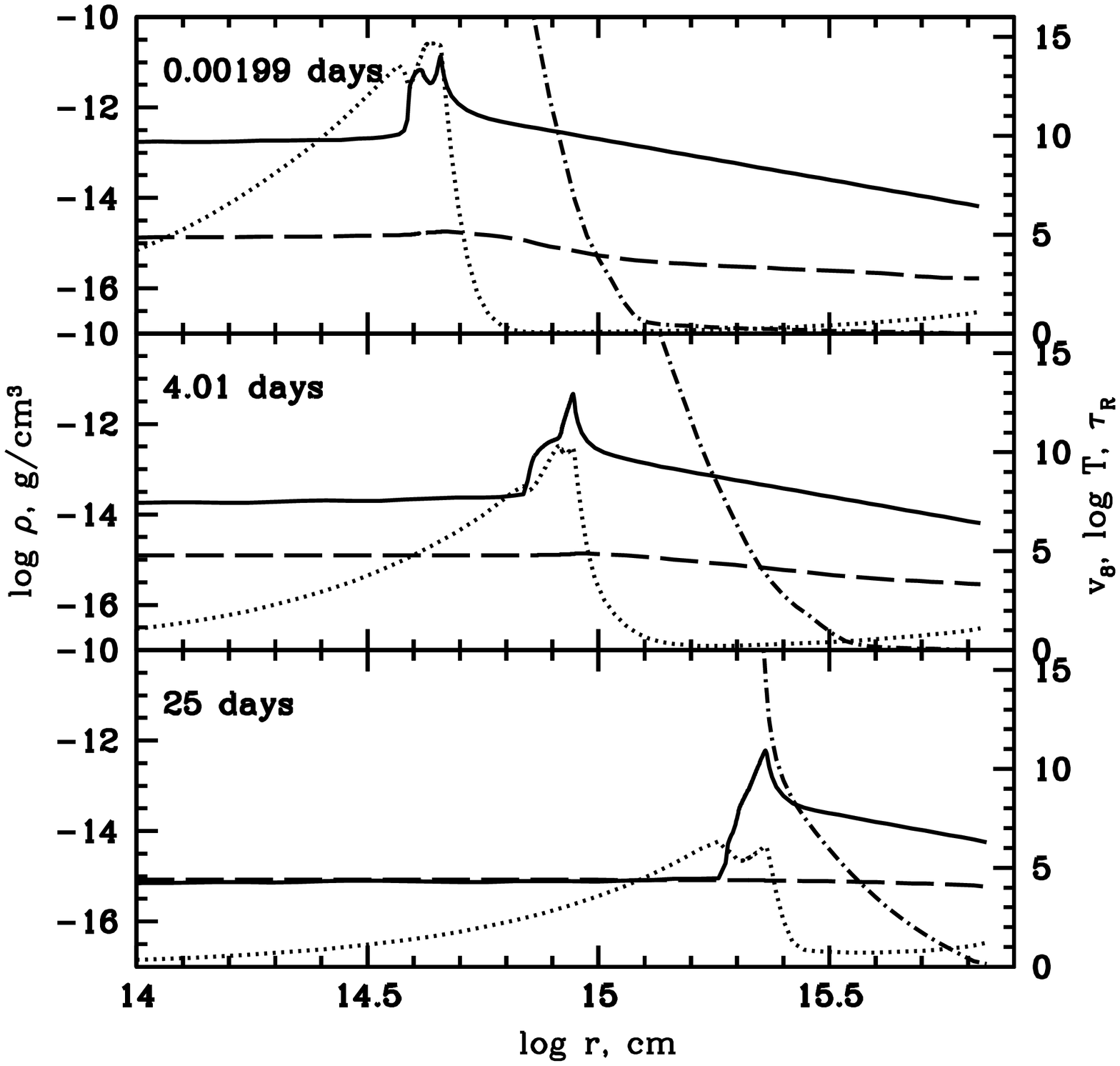}\\%\hfill
\includegraphics[width=0.74\linewidth]{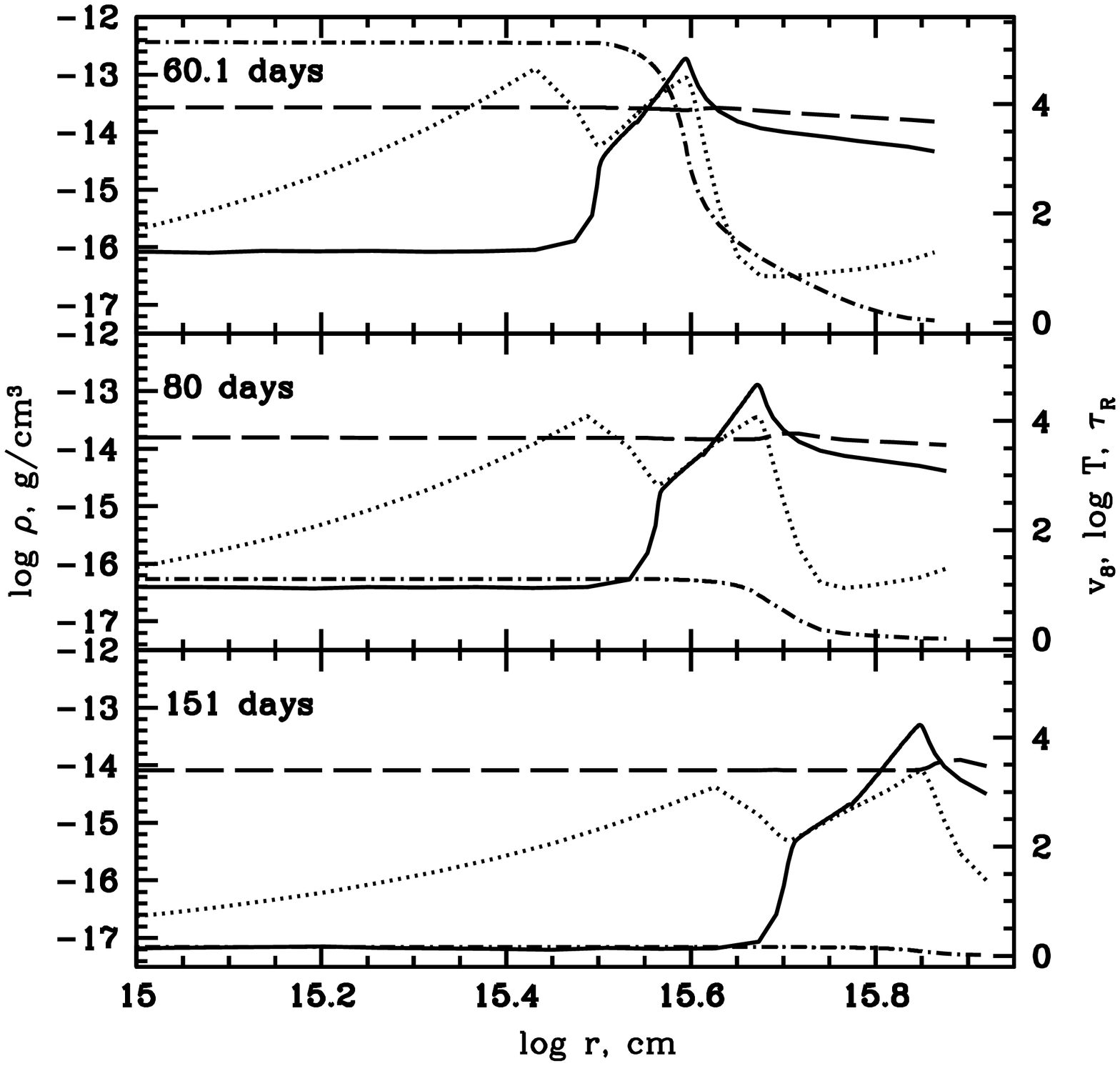}
\caption{
Evolution of radial profiles of the density ({\it solid lines}),
velocity (in $10^8$~cm~s$^{-1}$, {\it dots}), gas temperature ({\it dashes}),
and Rosseland optical depth ({\it dash-dots})
for the model N0 from \citet{sorokina2016}.
The scale for the density is on the left Y axis,
for all other quantities,  on the right Y axis.
Upper panel: pre-maximum hydrodynamical structure for three moments, very soon after the explosion and at days 4 and 25.
Lower panel: the same parameters, but after maximum, at days 60, 80, and 151.
Note that different scales for the axes are used on the upper and lower panels.
        }
\label{fig:hydroSLSN}    
\end{figure}

\subsection{Hydrodynamical properties} \label{subsec:InteractionHydro}
{\it Shock propagation through the optically thick extended CSM}. 
The shock waves produced at the ejecta-CSM interaction are initially radiation-dom\-i\-nat\-ed.
The photons behind the shock bear a large amount of energy compared to  the gas energy.
The photons are locked within the CSM and heat it up making the shock front wider.
Precursor heating extends to the optical depth $\tau=c/v_s$ from the shock front,
where $v_s$ is a shock velocity.
When it reaches the outer edge of the CSM, the photons that were heated on the shock start to gradually leave the system.
The shock begins to break out.
Due to the large size of the system the breakout stage lasts much longer than it does for standard SN~Ib/c.
In the latter case, $R<10\Rsun$ and the shock breaks out for less than $R/v_s\sim 200$~s,
while in the case of CSM-interacting SLSN with typical $R\sim 10^5\Rsun$, the shock breakout time is proportionally larger and lasts a few months.
On the whole the light curve of the SLSNe in the CSM-interaction scenario can be considered as a prolonged shock breakout.
At some stage the shock can become strongly radiative.
The temperature jump at the shock almost disappears, and the shock becomes almost isothermal.
In this case, the density can jump up by a few orders of magnitude,
and a rather dense and thin layer forms behind the shock.
A very detailed study of the structure of the SN shocks is provided by \citet{Weaver1976}.
More details are presented in \citet{tolstov2015}.
The basic formulas can be found in \citet{blinnikov2016}.

Fig.~\ref{fig:hydroSLSN} shows the results of a numerical calculation of the ejecta-CSM interaction for one of the SLSN models
that demonstrates how the radiative shock propagates throughout the dense extended CSM 
and what happens with the Rosseland optical depth (and therefore, photospheric radius) at the same time.
While the CS envelope is cool before the SN explosion, it is transparent to the radiation,
but when it becomes warm due to the pre-heating by the photons from the shock,
 it quickly becomes optically thick and strongly diffusive.
The photosphere ($\tau_{\rm R}\sim 1$) shifts to the outer edge of the CSM.
As a result, the shock is left  deep inside the photosphere for a few months.
%This defines the possibility of the measuring of the X-rays from SLSNe.
Conditions are not appropriate for hard emission at the early stages despite  the high velocity of the shock.
But an ultraviolet flash, or perhaps even an X-ray flash, can appear later,
when the shock becomes weaker, closer to the outer edge of the CS envelope.
This can happen in the fading stage of the light curve evolution.
SCP~06F6 is a possible example of such a SLSN originating from interaction
due to the X-ray flash observed about 2 months after the maximum \citep{GaensickeXray06F6}.
At the earlier stage, near  maximum, the temperature conditions are appropriate for  strong emission in the near UV range.
Almost all SLSNe are relatively blue near maximum,
and the shock heating in the CSM interaction scenario reproduces the color better then other mechanisms \citep{tolstov2017}.

\subsection{Light curves} \label{subsec:InteractionLCs}
{\it Analytical models.} % written by TM
Analytical methods to obtain rough estimates for the interaction-powered SNe are available. In this review, we focus more on numerical studies and we only briefly mention the analytic studies.

The dense CSM required to explain the large luminosity of SLSNe by the interaction is optically thick and we need to take large optical depths of the dense CSM into account in treating them analytically unlike the case of less luminous Type~IIn SNe \citep[e.g.,][]{chevalier2003,moriya2013iinana}. When the CSM optical depth is large enough, the shock breakout itself can occur in the dense CSM. \citet{chevalier2011irwin} provide a simple analytic way to estimate the rise time and peak luminosities followed by the shock breakout in the dense CSM. They show that the LC features differ depending on where the shock breakout occurs in the dense CSM. This method is extended to the case of non-steady mass loss by \citet{moriya2012tominaga}.

\citet{chatzopoulos2012semiana,chatzopoulos2013semiana} develop a semi-analytic approach to estimate the overall LC properties of the interaction-powered SNe that are commonly used recently. Although it is a simple powerful method to estimate the CSM properties, it is important to keep its limitation in mind. The semi-analytic method is based on the formalism of \citet{arnett1980} that assumes that the heating source is at the center of SNe. However, the energy source of the interaction model is not located at the center.
It moves outwards as the shock progresses. 
Moreover, it can be situated within a very thin layer, since gas between the forward and the reverse shocks can cool down strongly.
In this case, it is almost impossible to spatially separate the emission from the two shocks,
while the analytic solution by \citet{chatzopoulos2012semiana,chatzopoulos2013semiana} takes 
the radiation of the forward and the reverse shocks into account separately,
which leads to an inaccurate luminosity prediction.
Indeed, the CSM configuration obtained by the semi-analytic model has been shown to be inconsistent with those found by numerical approaches \citep{moriya2013sn2006gy,sorokina2016}. \citet{chatzopoulos2013semiana} study how to find the CSM configuration from the semi-analytic model that better matches the numerical results.

{\it Numerical modeling.} 
Numerical modeling allows  taking  more physical processes into account and simulating them in detail,
so the results should be more robust.

The interaction scenario is challenging for  numerical modeling.
A numerical code must be able to resolve optically thick shocks.
This is a reason why the number of SLSN interaction models calculated up to now is not very large.
The very first numerical modeling of the SN ejecta interaction with the dense extended medium has been done by \citet{Grasberg1971,FalkArnett1977,GrasNad1987}.
More recently, most calculations were produced with the radiation hydrodynamic code {\sc stella} \citep{blinnikov1998stella,blinn2006stella}.

{\it SLSNe II.}
Hydrogen-rich SLSNe are spectroscopically similar to SNe~IIn,
and therefore, their origin from the interaction of SN ejecta with slowly expanding dense CSM is the most probable scenario.
To provide the high luminosity of SLSN, the CSM is expected to be sufficiently more massive than for SN~II case.

\citet{woo2007PPINatur} suggest a pulsational pair instability (PPI) scenario for the last stage of the pre-SN evolution,
which leads to the formation of the required density structure.
PPI can happen for  stars with an initial main-sequence mass 95--130~M$_\odot$, 
or, more certain, with a helium-core mass 40--60~M$_\odot$.
Near the end of their lives,
when the central temperature exceeds $10^9$~K, 
such stars lose  stability through  electron-positron pair creation.
Explosive oxygen burning in the cores of these stars cannot unbind the whole star,
but leads to the ejection of a very massive envelope with  large kinetic energy.
The remaining stellar core contracts,
and some time later, 
when the central temperature again becomes high enough for  pair creation,
another ejection happens.
The process repeats until the stellar mass becomes too low for the pair instability,
and the core finally collapses due to iron dissociation.

As a result of the PPI scenario, the final core collapse happens inside the system of the massive expanding shells.
Collision of these shells, as well as of the final ejecta with shells,
can lead to a very bright event, as observed for the SLSNe.

\citet{woo2017PPI} presents a wider study of the models of the stellar evolution exploding as PPISNe.
He concludes that no solar metallicity star ends up as PPISN since it loses too much mass through the wind,
while stars with the metallicity $Z/Z_\odot < 0.1$ can lead to this end.
For SLSNe~II, whose host galaxies have metallicities in a wide range, including  high-metallicity objects \citep{perley2016slsnhost}, 
this means that the PPI explosion mechanism is not so likely to explain all these events, though more study of PPI is still required.
At the same time, PPI may be promising to explain SLSNe~I, since they prefer  low-metallicity galaxies 
and their photospheric velocities are much larger than what more standard stellar mass loss mechanism can provide.

\citet{woo2007PPINatur} apply the PPI scenario to one of the first SLSN~II, namely, SN~2006gy.
They found that the collisions of two subsequent mass ejections of about 25~M$_\odot$ and 5~M$_\odot$,
separated in time by less than 7 years,
with a total energy of about $3 \times 10^{51}$~ergs,
provides a good fit to the observed light curve of SN~2006gy.

\citet{moriya2013sn2006gy} studied a large number of toy models for the colliding ejections 
and found slightly different parameters for the best-fit model for SN~2006gy.
Both ejections contain a mass of about 15~M$_\odot$ (the second one could be lighter), 
and their kinetic energy is $4 \times 10^{51}$~ergs, which is only a little larger than the energy of normal SNe.
The influence of the density distribution on the light curve is also studied,
which showed that the first ejection could not be formed by a steady wind.

Despite a slightly different estimate of the ejection parameters,
both \citet{woo2007PPINatur} and \citet{moriya2013sn2006gy} converge on the conclusion that two subsequent ejections 
of a mass of a few tens solar mass with an energy close to the typical one for  SN explosions within a few years 
can easily lead to the very high luminosities typical for SLSNe.
%% added by TM %%
Similar results are obtained by \citet{dessart2015} and \citet{vlasis2016}, who also investigate aspherical CSM configurations.

\begin{figure}
\centering
\includegraphics[width=0.48\linewidth]{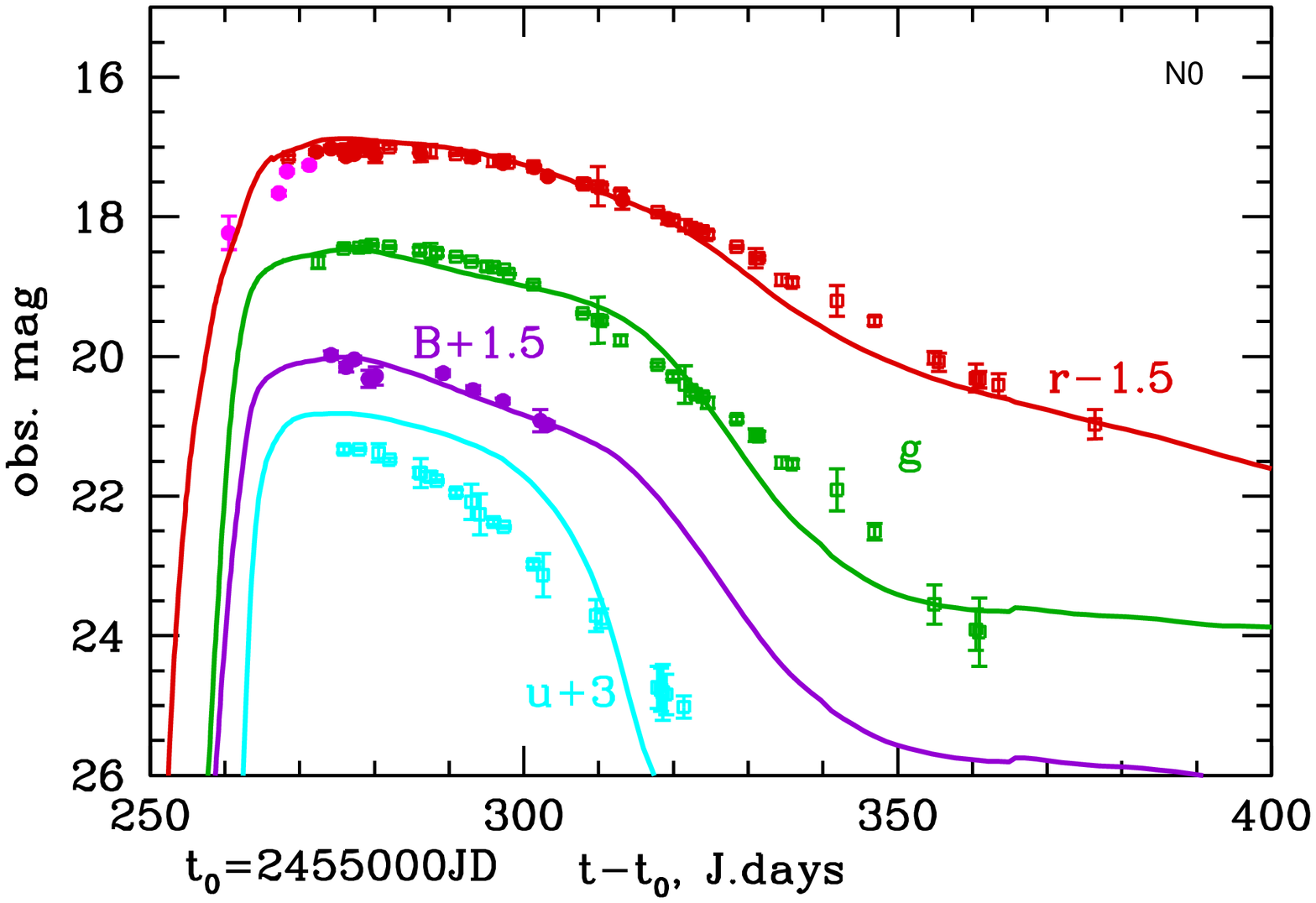}\hfill
\includegraphics[width=0.45\linewidth]{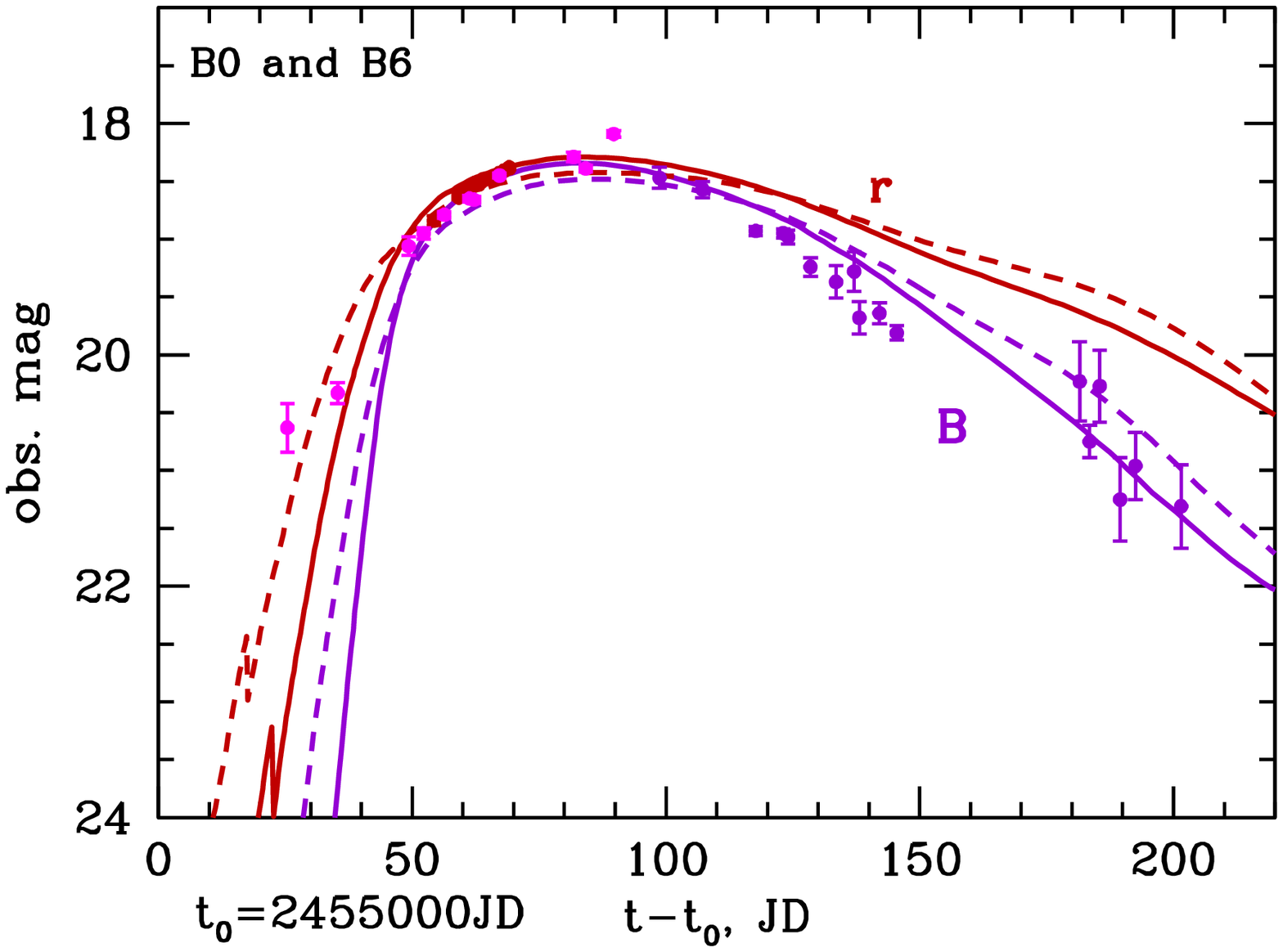}
\caption{The narrow light curve of SN~2010gx ({\it left}, \citealt{pastorello2010sn2010gx}) and the broad one of PTF09cnd ({\it right}, \citealt{quimby2011slsnic}) in different filters.
Observations are shown with {\it dots}, calculations \citep{sorokina2016} with {\it lines}. 
{\it Solid} and {\it dashed} lines in the right plot mean different composition:{\it solid lines} correspond to the model B0 with the outer envelope
                   containing 90\% carbon and 10\% oxygen,
                   {\it dashed lines} correspond to the model B6 similar to B0,
                   but roughly half of C and O in the outer envelope is replaced with He.
        }
\label{fig:fitLCinteract}    
\end{figure}

{\it SLSNe I.}
In the modeling of hydrogen-poor SLSNe, one more parameter appears in addition to those that are important for hydrogen-rich events.
It is the chemical composition of the ejecta and the CSM.
There are several observational hints for choosing the correct composition for numerical modeling of SLSN~I light curves and spectra.
Observationally, strong broad oxygen lines are typical for SLSNe~I from the very beginning, 
but very few SLSNe~I have helium lines in their nebular spectra.
So the observations favor a CO composition for the ejecta and CSM.
Numerical modeling of the multicolor broad-band light curves \citep{sorokina2016} is in agreement with the observations from this point of view.
The presence of helium  leads the light curve to rise much longer than observed,
while a CO composition results in a faster rise. 
This difference is explained by different opacities for CO and He.
CO material requires a lower temperature for excitation and ionization, so the carbon-oxygen CSM becomes opaque 
soon after the shock starts to heat the CSM gas in front of it.
Helium needs a higher temperature to become opaque, and it takes a longer time.
Some amount of helium can be mixed with the CO gas.
The light curve is not much changed when carbon and oxygen are the main absorbers.
The photometric colors near maximum help determine the proportions of C to O,
since these elements have different spectral line distributions.
Oxygen is more absorptive in the blue, and it shifts the larger part of the emission to this wavelength range.
The modeling by \citet{sorokina2016} shows that models with a C to O ratio from 0.7 to 0.9 
show the best fit to the observations of the narrow and the broad SLSN I light curves, respectively.
The light curves are shown in Fig.~\ref{fig:fitLCinteract},
and demonstrate how the pure ejecta--CSM interaction scenario, without any radioactive material, can reproduce the emission of SLSNe~I.
The masses and energies of the models are similar to those needed for SLSNe~II.
A fast evolving SLSN~I can be reproduced by the interaction with about 10~M$_\odot$ CS envelope,
while for slowly fading events, a few tens M$_\odot$ of hydrogen-poor material is needed.
The explosion energy in any case is a few $10^{51}$~erg.

{\it Bumpy light curves.} Light curves of some SLSNe~I show some undulations on the fading part \citep{inserra2017slsnslowlcprop}.
The CSM interaction scenario explains this in the most natural way:
each bump on the light curve can denote the collision of CSM layers or the SN ejecta with a CSM layer.
If the progenitor star experiences several episodes of massive ejection before the final collapse 
as it can happen in the PPI mechanism,
then several collisions of CSM layers could be later observed as bumps on the SLSN light curve.
In the spectra of the SLSNe~I with the most undulating light curves,
hydrogen lines appear at a late stage \citep{yan2017slsniclateh}.
This also can be considered as a proof for the CSM interaction scenario for some SLSNe~I.
Initially the luminosity of the SLSN is explained by the collision of the innermost hydrogen-free layers,
while the outer layers that can contain some hydrogen are cold and transparent.
Later, the inner hydrogen-poor layers, that are faster, reach the hydrogen-rich one,
heat it, and the hydrogen lines appear in the spectrum.

The origin of the early pre-maximum luminosity excess that is observed in many  SLSNe~I is more questionable.
\citet{moriya2012} qualitatively explain the possible origin of this excess by heating of a detached CS envelope.
Its opacity is enhanced after ionization leading to a temporary decrease in luminosity before a rise to the main maximum.
More detailed numerical modeling is needed for this mechanism.
Some other explanations of the pre-maximum excess are described in Sect.~\ref{sec:comparison}.

\subsection{ Pros and cons for the interaction scenario }\label{subsec:InteractionProsCons}
Summarizing the  above material, we point out the strengths and weaknesses of the CSM interaction model for SLSNe.

The interaction scenario for SLSNe is favored due to the following reasons:
\begin{itemize}
 \item SLSNe~II are spectrally very similar to SNe~IIn that are explained by  CSM-ejecta interaction.
    For a more massive CSM, a more luminous (and superluminous) SN is expected.
 \item SLSN~I light curves with a wide range of rising and fading timescales can quite easily be modeled by the interaction with a CSM of different mass.
The larger the light curve timescale, the more massive CSM is needed. Only the widest light curves are problematic for  modeling with interaction.
 \item The undulation on the fading part of the light curve is most naturally explained by a sequence of collisions of several CSM envelopes or clouds.
 \item  Measuring masses of different elements from the nebular spectra of SLSNe \citep{jerkstrand2017slsnnebular}, energetics of these events, and other observational properties show that 
 SLSNe originate from massive stars.
     Mass loss is typical for these objects \citep{VeryMassiveStars2015}.
    In this sense, some kind of CSM interaction is unavoidable after the SN explosion.
    However, the amount of such material must be very large to provide the strong interaction that is required to explain the SLSN phenomenon.
\end{itemize}

The last item from the list above takes us to the list of difficulties
that still remain for the interaction mechanism.
\begin{itemize}
 \item At least a few solar masses must be lost by a SLSN progenitor during a few months before the explosion 
    to explain the brightness and the duration of the SLSN events.
    Although this is not impossible from the stellar evolution point of view, there is no detailed understanding of the very last stages of stellar life yet, neither theoretical, nor observational. The required CSM mass
    is comparable to observed mass-loss rates only from  luminous blue variables (LBVs) in outburst \citep{Smith2011MNRAS}.
    For the metal-poor stars that are able to preserve considerable mass without losing it through a steady wind till the end of their evolution, 
    eruptive mass loss with the ejection of several solar masses of gas prior to the explosion could be very important \citep{SmithOwocki2006}.
    Anyway, a detailed mechanism for the LBV eruptions is still unknown, 
    as well as the details of the mass loss by the PPI mechanism.
    It is not, of course, an argument against the origin of SLSNe from CSM interaction, but
    this topic still requires both theoretical and observational efforts.
 \item High photospheric velocities (10,000 -- 20,000~km/s) are measured in SLSNe~I for about a month from maximum.
    Existing numerical interaction models do not reproduce these velocities.
    But an explosion energy only a little larger than standard has been considered in  modeling.
    Enhancing the explosion energy up to a few $10^{52}$~erg and the kinetic energy of the CSM expansion to about  $10^{51}$~erg
    can improve the situation.
    However, this energetic model would bring another problem:
    such energetic explosions must be separated in time longer than needed for getting a strong interaction because the stronger an eruptive mass ejection is, the longer time is needed for the star to reach the instability conditions for the next explosion or eruption. For example, an estimate of the time intervals between the eruptions for the PPI model can be found in \citet{woo2017PPI}.
 
 \item The slowest (i.e., widest) observed SLSN~I light curves together with high velocities are extremely difficult to explain solely by CSM interaction.
    In this case a combination of energy sources, like CSM interaction for the maximum plus radioactivity for the tail similar to what has been done in \citet{tolstov12dam2017,tolstov2017}, is more promising. 
\end{itemize}

Detailed numerical modeling for an extended set of explosion parameters is still needed to understand the applicability of the CSM interaction model to the SLSNe,
especially for the hydrogen-poor ones.
Not only the light curves must be fitted, but also the spectra.
PPI seems to be the best mechanism for making a CSM density and velocity structure that would lead to a SLSN event after the interaction with the SN ejecta.
But PPI itself  must be better studied in order to understand the limitations it places on the CSM structure.

\section{Magnetar-powered models}\label{sec:magnetars}
The view that pulsars could play a role in powering supernovae was proposed soon after the discovery of pulsars.  \cite{ostriker1971} proposed that the spin-down of a central pulsar could power both the supernova explosion and the light from the supernova.
In this model, the pulsar creates a bubble of relativistic fluid that sweeps up and then accelerates the surrounding star.
However, this model was not found to be in accord  with supernova observations.
The current prevailing model is that the  power for the supernova is injected over a short period of time (seconds) and the 
pulsar bubble then evolves in the freely expanding ejecta set up by the initial explosion.
This model was proposed for the Crab Nebula and other young pulsar wind nebulae \citep{chevalier1977} and for the light from
supernovae \citep{gaffet1977}.
Generally, the light from core collapse supernovae does not require power input from a pulsar, and can be explained by a combination
of initial explosion energy, radioactivity, and possible circumstellar interaction.
However, the supernova SN 2005bf showed two broad peaks, implying the operation of some additional mechanism in addition to the usual.
\cite{maeda2007} suggested that power from pulsar spin-down could be the source of the additional power.
The observed luminosity and timescale required a highly magnetized pulsar ($> 10^{14}$ G), or magnetar, with an initial spin period in the ms range.
\cite{kasen2010} and \cite{woosley2010} found that similar parameters could explain the timescale and luminosity of SLSNe, and this has
become one of the leading contenders for the light from SLSNe.
Here we discuss the basic magnetar model, giving some of the uncertainties of the model.
We primarily deal with simplified models that illustrate the physical principles.

The initial central explosion is assumed to occur on a timescale of seconds, followed by shock traversal of the star and the ejecta transition to free expansion on a
timescale of the doubling of the radius.
At that point, the velocity profile is given by $v=r/t$.
The density profile for the expanding gas can be approximated as a steep power law with radius at large radii and a flat power law at small radii
\citep{chevalier1989,matzner1999}.
We have the inner power law $\rho_{in}\propto t^{-3} (r/t)^{-\delta}$ and outer $\rho_{out}\propto t^{-3} (r/t)^{-n}$.
Here, we take $\delta = 0$.
The transition velocity between the inner and outer density profiles is 
\begin{equation}
%v_{tr}= \left[\frac{2(5-\delta)(n-5)}{(3-\delta)(n-3)}\frac{E_s}{M_e}\right]^{1/2}.
v_{tr}= \left[\frac{10(n-5)}{3(n-3)}\frac{E_s}{M_{ej}}\right]^{1/2}.
\end{equation}
For  $n=7$, we have $v_{tr}=4080 E_{51}^{1/2}M_5^{-1/2}\kms$, 
where $E_{51}$ is $E_s$ in units of $10^{51}$ ergs
and $M_5$ is $M_{ej}$ in units of $5\Msun$,
where $E_s$ is the supernova energy and $M_{ej}$ is the ejecta mass.
The value of $E_s$ is generally taken to be $1\times 10^{51}$ ergs, which is a value that is obtained for cases where the energy is well determined.

The newly formed rapidly rotating neutron star is assumed to drive a supersonic/superAlfvenic wind, powered by the spin-down of the neutron star.
The details of how the transition is made from an accreting neutron star to a pulsar wind are not known.
%A common assumption on the spin-down power is that the vacuum dipole expression applies.
A plausible estimate for the spin-down power is the result for a force-free magnetic field
\citep{spitkovsky2006}
\begin{equation}
\dot E\approx \frac{\mu^2 \Omega^4}{c^3}\left(1+\sin^2 \alpha\right),
\end{equation}
where $\mu = B R^3$ is the magnetic dipole moment, $\Omega=2\pi /P$ is the spin frequency,  $\alpha$ is the angle between the
magnetic and rotation axes, $B$ is the dipole field at the neutron star surface, and $R$ is the neutron star radius.
Here, $R$ is taken to be 10 km.
The neutron star rotational energy is $E=\frac{1}{2}I\Omega^2$ where $I$ is the neutron star moment of inertia, taken to
be $1\times 10^{45}$ gm cm$^2$.
The initial spin-down time is
\begin{equation}
t_{sd}=\frac{E_0}{\dot E_0}  = 0.14 \left(\frac{P_0}{\rm ms}\right)^{2}\left(\frac{B_p}{10^{14}{\rm G}}\right)^{-2}{\rm~day},
\end{equation}
where the subscript $0$ refers to the time $t=0$.
The evolution of the spin-down power is given by
\begin{equation}
\dot E = \dot E_0 \left(1+\frac{t}{t_{sd}}\right)^{-2}.
\end{equation}

The structure of pulsar magnetospheres is not a fully solved problem; there are other possibilities for the spin-down.
For the vacuum magnetic dipole case
\begin{equation}
\dot E_{vac} = \frac{2\mu^2 \Omega^4}{3c^3}\sin^2 \alpha.
\end{equation}
In the force free case, there is spin-down for an aligned rotator, but not in the vacuum case.

%More generally, the evolution of the spin-down can be expressed in terms of the evolution of the forque $\dot \Omega =-f\Omega^m$, where $m$ is the braking index and $f$ is typically taken to be a constant.
More generally, the evolution of the spin of a pulsar is often taken to be of the form
\begin{equation}
\dot \Omega \propto -\Omega^m
\label{brake}
\end{equation}
where $m$ is the braking index.
The value of $m$ for the magnetic dipole case is 3 and the constant of proportionality  in that case can be related to the neutron star moment of inertia and magnetic
field, as discussed above.
The value of $m$ can be determined from observations by $m=\ddot \Omega \Omega/ \dot\Omega^2$.
%Observations of pulsars with ages $\sim 1,000$ yr do not have a spin-down that is closely aligned with the vacuum dipole value.
%The torque on the neutron star can be written as
The power from the spin-down with constant $m$ is
\begin{equation}
\dot E = \dot E_0 \left(1+\frac{t}{\tau}\right)^{-(m+1)/(m-1)},
\end{equation}
where $\tau$ is a spindown timescale that cannot now be written simply in terms of the magnetic field.
Measured values of $m$  for ordinary pulsars are $<3$. 
For PSR J1119--6127, $m=2.91$, close to the magnetic dipole value.
Other values are lower; the long term value is $m=2.34$ for the Crab Pulsar and 1.7 for the Vela pulsar \citep{espinoza2017}.
Possible reasons for deviations from the magnetic dipole value are evolution of the neutron star moment of inertia or magnetic field,
or a more complex magnetic configuration than a dipole.
These considerations show that pulsars generally do not follow the magnetic dipole spin-down and the late evolution of $\dot E$ (at $t\gg \tau $) is
steeper than in the force-free dipole case; for the case of the Vela pulsar, $\dot E\propto t^{-3.9} $ at late times.
The parameters for the rapidly rotating magnetars of interest here are quite different from the ordinary pulsars, but these objects show that
pulsar spin-down can deviate from the force-free dipole case.

%The power input from the pulsar is unlikely to be spherically symmetric.
%The angular dependence of the power is $\sin^2 \theta$, where $\theta$ is the angle between the magnetic axis and the rotation axis.
%Camus have considered the effect of this dependence in 2-D models of pulsar bubble evolution.

We next consider the evolution of the pulsar bubble on the assumption of adiabatic motion.
In the case of a pulsar wind nebula, the shocked pulsar wind of relativistic particles and magnetic field can be approximated by
a $\gamma =4/3$ fluid, where $\gamma$ is the adiabatic index.
For the relatively young SLSN case, the assumption is that the pulsar power thermalizes so the radiation field is the dominant 
energy density.  
Again, an initial value $\gamma =4/3$ is appropriate.
If spherical symmetry is assumed, the expansion of the pulsar bubble can be found in the thin shell approximation.
Even if the outer shock wave is not radiative, the swept up shell is thin.
%The shell is accelerated into the freely expanding ejecta with radius
%For  $n=7$, we have $v_t=4080\, E_{51}^{1/2}M_5^{-1/2}\kms$,  where $E_{51}$ is $E_s$ in units of $10^{51}$ ergs and $M_5$ is $M_e$ in units of $5\Msun$.
If the pulsar provides a steady power $\dot E$, the radius of the swept up shell is \citep{chevalier1977,blondin2001}
\begin{equation}
R_s=1.50 \left(\frac{n}{n-5}\right)^{1/5} \left(\frac{n-5}{n-3}\right)^{1/2}\left(\frac {E_{s}^3\dot E^2}{M_{ej}^5}\right)^{1/10} t^{6/5},
\label{radius}
\end{equation}
where the thin shell approximation has been made.
We have 
\begin{equation}
t_{tr}=\frac{66(n-5)}{25 n}\frac{E_s}{\dot E},
\end{equation}
where $t_{tr}$ is the time for the shell to reach the
inflection point in the supernova density profile.
%%change
Approximating $\dot E =E /t_{sd}$ for a time $t_{sd}$,    %%and $\dot E_p =0$ thereafter
 we have
\begin{equation}
t_{tr} = 0.19 B_{14}^{-2}P_{ms}^4 E_{51}{\rm~day}. 
\end{equation}

Once the shell has traveled through the flat part of the supernova density profile, it comes into the steep power law part of the density profile.
The mass added from the outer ejecta is negligible compared to that in the shell and, for 1-dimensional (1-D) expansion, the evolution of the shell radius goes to $\propto t^{1.5}$
as in the solution of \cite{ostriker1971}.
This is another self-similar solution.

The acceleration of the shell by the hot bubble is subject to Rayleigh-Taylor instability (RTI).
The action of the RTI while the shell is in the flat part of the ejecta density profile has been well studied
\citep{porth2014,bucciantini2004}.
Despite the growth of instabilities, the instability saturates and the flow remains self-similar; the outer shock front has a radius that is 7\% larger than in the 1-D solution \citep{blondin2017}.
Most of the swept up mass is in the shell, although there is also some matter that is in fingers that extend in towards the neutron star.
The RT fingers have relatively little effect on the termination shock of the pulsar wind.
Global simulations including a magnetic field show qualitatively similar properties, although the Kelvin-Helmholtz instability is suppressed on small scales.

Once the shell reaches the transition point in the density profile, there is a qualitative change in the evolution.
The preshock ejecta are unable to contain the pressure of the shocked wind bubble \citep{chevalier2005} and the bubble matter blows out through the
shell \citep{chen2016,suzuki2017,blondin2017}.
Channels form in the swept up ejecta through which matter from the inner wind bubble flows.
\cite{suzuki2017} used a special relativistic code, allowing for a relativistic pulsar wind that shocked and slowed on passing through the termination shock.
When the gas was able to escape through channels, relativistic velocities were again attained.

The power provided by the magnetar initially comes out in a relativistic wind of particles and Poynting flux.
%%  something on thermalization 
Assumptions must be made about the fraction of the wind power that goes into thermal radiation field.
There might also be some fraction that remains in nonthermal form.
The mechanisms giving rise to thermalization are not well understood.

Standard models for calculating the light curves of SLSNe I \citep{chatzopoulos2012semiana,inserra2013firstmagnetar,nicholl2017} have been proposed based on the formalism developed
by \cite{arnett1980,arnett1982} for ordinary supernovae.
The model assumes that the supernova gas is homologously expanding and is characterized by a constant opacity $\kappa$ for the thermal radiation.
The emitted luminosity is
\begin{equation}
L(t)=2 e^{-(t/t_d)^2}\int^t_0 e^{(t^{\prime}/t_d)^2}\frac{t^{\prime}}{t_d} \dot E(t^{\prime})\frac{dt^{\prime}}{t_d}.
\label{lumin}
\end{equation}
where
\begin{equation}
t_d=\left(\frac{10\kappa M_{ej}}{3\beta v_{ej}c}\right)^{1/2}
\end{equation}
is a diffusion time that gives a characteristic timescale for the light curve, 
%%$\kappa$ is the opacity for the thermal radiation, 
$v_{ej}$ is the velocity at the outer edge of the ejecta, and $c$ is the speed of light.
Here, $\beta$ is a constant that depends on the density distribution of the freely expanding ejecta.
We have $\rho_{ej}\propto \exp(1.723 Ax)$, where $A$ is a constant and $x=r/v_{ej}t$.
For uniform density ejecta ($A=0$), $\beta =13.8$ and for $A=-1$, $\beta=13.4$; $\beta =13.8$ is typically chosen in magnetar models.
As discussed above, a fairly flat density distribution is a reasonable approximation for the
freely expanding ejecta in a core collapse supernova.
However, in the present case, the density distribution is strongly affected by the magnetar wind nebula, causing some uncertainty in the model fits.

There is the assumption in equation (\ref{lumin}) that the initial stellar radius is much smaller than the supernova radius so that terms involving
the initial radius and thermal energy do not appear.
The initial explosion energy is lost to adiabatic expansion.
Another assumption in equation (\ref{lumin}) is that all the  pulsar power goes into the thermal radiation, but it may only be some fraction $f$
that is thermalized.
There have been various suggestions on how nonthermal effects can be included.
\cite{kotera2013} suggested that a constant fraction of the pulsar power goes into the thermal radiation.
However, \cite{wang2015} noted that models using equation (\ref{lumin}) fit the observations at early times, but fall below the observations at late times.
They suggested that
\begin{equation}
f=1-e^{Bt^{-2}},
\end{equation}
where
\begin{equation}
B=\frac{3 \kappa_{\gamma}M_{ej}}{4\pi v_{ej}^2}
\end{equation}
and $\kappa_{\gamma}$ is the opacity for high energy photons.
This expression has the property that at early times, nonthermal photons are completely thermalized, and at
late times there is optically thin absorption of photons, assuming that the absorbing material is uniform.

\begin{figure}
  \includegraphics[width=\textwidth]{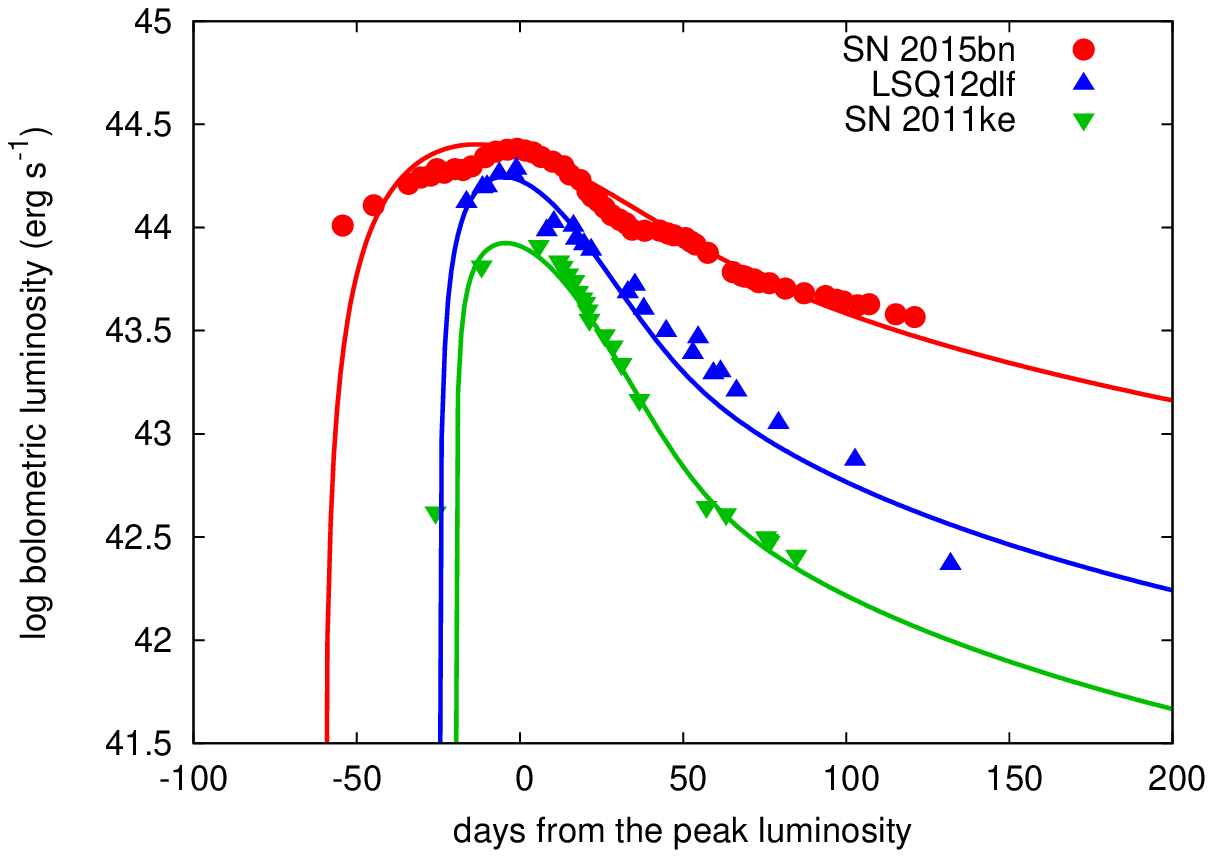} 
\caption{
Examples of magnetar-powered LCs fitted to SLSNe with different LC evolution. The SLSN bolometric LCs are obtained from \citet{nicholl2016sn2015bnearly}. The model LCs are calculated with a semi-analytic way based on \citet{arnett1982} as in \citet{inserra2013firstmagnetar}. The magnetar spin down energy is assumed to be thermalized with 100\% efficiency. The magnetic field strengths and the spin periods in the models are: $10^{14}~\mathrm{G}$ and 2~ms (SN~2015bn), $3.5\times 10^{14}~\mathrm{G}$ and 2.5~ms (LSQ12dlf), and $7\times 10^{14}~\mathrm{G}$ and 2.5~ms (SN 2011ke).
}
\label{fig:magnetar_lightcurve}       
\end{figure}

This model, or some variation on it, has been widely used in modeling the light curves of SLSNe I.
Different researchers make different assumption, hindering the comparison of results.
\cite{nicholl2017} have analyzed the light curves of 38 SLSNe I in a consistent way, using the 
radiative transfer approximation of Arnett and the energetic photon escape of \cite{wang2015}.
They assumed the vacuum magnetic dipole spindown for the pulsar power and made a
blackbody assumption to obtain multi-color light curves for the supernovae.
A result of their model fitting is shown in Fig.~\ref{fig:magnetar_lightcurve}.
The models fit the cases with a decline that mimics that of $^{56}$Co, like SN 2007bi.
They deduce the following ranges (1-sigma) of parameters from their models: initial period  $1.2-4$ ms, $B=(0.2-1.8)\times 10^{14}$ G, and $M_{ej}=(2.2-12.9) ~\msun$.
Using similar methods, \cite{liu2017} had previously modeled 19 supernovae and obtained good fits for 14 objects.
They found parameter ranges: initial period $1.2 - 8.3$ ms, $B=(0.2-8.8)\times 10^{14}$ G, and  $M_{ej}=1 - 27.6 ~\msun$.
These results give some idea of the range of parameters in present modeling. \citet{yu2017magfit} also modeled 31 SLSNe similarly and found parameter ranges of initial period $1.4-12$ ms, $B = (0.5-4.8) \times 10^{14}$ G, and $M_{ej} = 0.5-17.9~\msun$.

Most modeling efforts have concentrated on the supernova light curves, but there has also been attention to the spectra of SLSNe.
\cite{dessart2012magni} calculated steady-state, 1-D, non-LTE spectral models with power input from a magnetar.
They found that the magnetar has important effects when the energy deposited by the magnetar becomes comparable to
the supernova energy.
In this case, the spectrum is blue, in accord with observations, and does not have the line-blanketing that occurs when the
power is produced by radioactivity.
Broad lines can be produced, as observed, if the ejecta mass is relatively  small.

\cite{bersten2016} also went beyond the simple models in their light curve models, using a 1-D, special relativistic, radiation hydrodynamics code.  They found that hydrodynamical modeling is especially important when the deposited magnetar energy exceeds that of the initial supernova, consistent with the arguments given above.

Although the magnetar model has had considerable success in reproducing the light curves and spectra of SLSNe I, these results
follow from a certain power supplied over a certain time  and do not provide a ``smoking gun'' for the magnetar model.
There has been interest in whether some of the detailed features of SLSN I observations require a magnetar.
As noted in Section 2.2, some SLSNe I show a precursor in the light curve before the main peak.  
\cite{kasen2016} suggested that the precursor is emission from the breakout of the shock wave driven by the pulsar wind bubble.  However, the power generated by the shell sweeping into the freely expanding ejecta is small.  In order to obtain a significant precursor, \cite{kasen2016} suggest that only a fraction of the initial pulsar power is thermalized, reducing the main peak relative to the shock breakout.  Other suggestions for the precursor emission include circumstellar interaction \citep{moriya2012,piro2015} or a highly energetic supernova \citep{nicholl2015lsq14bdq}.

\cite{metzger2014} have discussed the breakout of an ionization front through the swept up shell.
They assume that a substantial fraction of the pulsar power goes into an ionizing spectrum radiated by energetic particles.
The ionization front is initially trapped in the shell but it is able to penetrate the shell as the column density of the shell
declines due to expansion.
This breakout can occur within months of optical maximum under certain conditions.
Accounting for the Rayleigh-Taylor instability of the shell would result in an earlier breakout if the pulsar wind evolves to this phase.
The ionization breakout theory was proposed in part to explain the luminous X-ray emission observed from SCP 06F6  \citep{levan2013slsnxray}.
The complete ionization of the shell allows the X-rays to escape.
However, after the detection with XMM, an observation with Chandra set a low upper limit on X-rays from SCP 06F6, and X-ray observations of other SLSN have
only yielded upper limits \citep{margutti2017slsnxray}.

There may be a connection between SLSNe and long gamma-ray bursts (LGRBs).
Both classes of objects occur in small, low metallicity galaxies.
An ultra-long GRB, GRB 111209A, has recently been observed along with a bright supernova, SN 2011kl \citep{greiner2015ulgrbslsn}.
\cite{margalit2017}  suggest that the two classes of objects have similar progenitors (rapidly rotating neutron stars) and that the alignment
between the spin axis and the magnetic axis is what determines the relative amount of jet vs thermal emission.
In this view, the aligned rotator case is expected to give strong jets.
The importance of jets in SLSNe is also discussed in, e.g., \citet{soker2017jet,gilkis2016jet}.

Although the magnetar model has attractive features, the case is not yet settled.
The multidimensional hydrodynamic models show that a complex 3-D structure results from power input from an energetic magnetar \citep{chen2016,suzuki2017,blondin2017}.
Models that account for that structure are ultimately needed.

We summarize pros and cons of the magnetar-powered models.\\
Pros:
\begin{itemize}
 \item The magnetar models can explain both the slowly declining and the rapidly declining light curves of SLSNe~Ic.
 \item  The observed velocities and temperatures are roughly reproduced.
\end{itemize}
Cons: 
\begin{itemize}
 \item The magnetar models have a number of free parameters.  There is not a ``smoking gun'' for the presence of magnetar power.
 \item The magnetar models do not naturally explain the bumps observed in the light curves of some SLSNe.
\end{itemize}

\section{Comparison}\label{sec:comparison}
Here we compare the three major suggested mechanisms discussed so far. SLSNe~IIn show clear signatures of CSM interaction and their LCs and spectra can be explained by the interaction-powered model (Section~\ref{sec:interaction}). We here focus on SLSNe~Ic whose luminosity sources are actively discussed now.

\subsection{Light curves}
Both interaction-powered and magnetar-powered models can provide reasonable fits to most SLSN~Ic LCs with a proper choice of the model parameters. The $^{56}$Ni powered models fail to explain the rapidly declining SLSN~Ic LCs as their late-phase luminosity evolution is governed by the nuclear decay rate of $^{56}$Co which is rather slow.

Some SLSN~Ic LCs show complicated structures that may prefer the interaction model. For example, the LCs of iPTF15esb \citep{yan2017slsniclateh} have two major peaks that are hard to  explain with a single luminosity source at the center and prefer the existence of a complicated dense CSM structure. The ``spiky'' bolometric LC of SN~2017egm is also suggests an interaction model \citep{wheeler2017sn2017egm}. The LC modulations observed in some SLSNe~Ic are also likely to indicate the existence of an outer energy source like a dense CSM. Even if there is a major central power source like a magnetar powering a SLSN~Ic LC, some dense CSM that has a minor contribution to the luminosity may also exist. A combination of  central and outer powering sources could better explain the general SLSN~Ic LCs.

Multi-frequency studies of SLSNe~Ic are helpful in distinguishing the different powering mechanisms.
For example, the magnetar model predicts a late-phase increase in X-rays at the ``ionization breakout" \citep{metzger2014}, while X-rays  in the interaction model are expected in some cases
after the shock breakout stage that lasts a few months. \citet{margutti2017slsnxray} report their observational efforts to detect SLSNe in X-rays but only a few detections are obtained so far and it is still hard to distinguish the powering mechanisms in this way. \citet{tolstov2017} recently suggested that the ultraviolet luminosity could be better explained by the interaction model. It is not totally clear yet if the ultraviolet luminosities are the decisive distinguisher among the models, given the uncertainties in how magnetars transform their spin energy to the surrounding ejecta. More studies are required in this regard. No $\gamma$-ray detections from SLSNe could indicate that the magnetic field strengths in the compact remnants of SLSNe are much less than required to power the LCs \citep{renault-tinacci2017gamma}.

The precursors found in many SLSNe~Ic are hard to explain with any model. If the precursor is from the shock breakout, the progenitor radii need to be more than $500~R_\odot$ with a reasonable assumption of the explosion energy \citep{nicholl2015lsq14bdq,piro2015}. These huge radii are unexpected in hydrogen-free progenitors and some mechanisms to make an extended envelope are required. The $^{56}$Ni-powered model only has the central heating source without much hydrodynamical effect and such extended envelopes are necessary to explain the precursor.

The CSM interaction and magnetar models are able to explain the precursor without invoking the extremely extended progenitor for SLSNe~Ic. For the CSM interaction model, \citet{moriya2012} illustrate that the precursor bump can be naturally made by the ionization change in the massive CSM to power the major LCs in SLSNe~Ic. Regardless of the luminosity source of the precursor, the precursor LC bump can be made by the massive CSM required to explain the main peak of SLSN~Ic LCs. In the magnetar-powered model, so-called ``magnetar shock breakout'' can occur in the ejecta, making a precursor on some occasions (\citealt{kasen2016}, see also Section~\ref{sec:magnetars}). When a huge magnetar rotational energy is released near the center of an exploding star, another shock wave can be launched. If the shock becomes strong enough to be a radiation-dominated shock, another shock breakout can occur when the shock reaches the surface of the ejecta resulting in a precursor bump observed in SLSNe~Ic. However, in 1D LC simulations, the shock breakout bump is usually not clearly seen as observed unless the central heating efficiency is artificially reduced \citep{kasen2016,moriya2016magbh}. More studies are needed to see if the magnetar shock breakout can generally explain the precursors.

\subsection{Spectra}
Spectra of SLSNe~Ic provides plenty of information to distinguish the powering mechanisms. The $^{56}$Ni-power model is generally less favored recently because of the lack of strong Fe absorptions and emissions in SLSNe~Ic (e.g., \citealt{jerkstrand2017slsnnebular}, see also Section~\ref{sec:56ni}). The models with a central engine like magnetars are generally found to match the spectra well \citep[e.g.,][]{mazzali2016slsnicsp,dessart2012magni}. SLSNe~Ic do not show narrow features that are found in less luminous Type~Ib \citep[e.g.,][]{pastorello2016typeibn} and Type~Ic \citep[e.g.,][]{ben-ami2014sn2010mb} SNe with  strong CSM interactions. There are some suggested mechanisms to hide the possible narrow lines in the interacting SNe \citep[e.g.,][]{chevalier2011irwin,moriya2012tominaga}, but they may not generally occur. Spectral modeling is required to judge if the lack of the narrow emissions in SLSNe~Ic is consistent with the interaction model.

\citet{chen2016lsq14mo} performed a detailed spectral modeling of the Type~Ic SLSN LSQ14mo. They show that its line features did not change for a while but the continuum significantly changed in the same period. The temperature deduced from the line features does not match that deduced from the continuum for some period. In their interpretation, the continuum may be affected by an external heating source like  CSM interaction in addition to a central heating source like a magnetar that is determining the temperature of the line forming region in the SN ejecta. This indicates that more than a single energy source is responsible for the LCs for all the periods. Different powering sources could be contributing at the same time to power SLSNe. Another example of this kind is Gaia16apd, which is suggested to be powered by a combination of CSM interaction and $^{56}$Ni heating \citep{tolstov2017}.  $^{56}$Ni heating may contribute to SLSNe~IIn in addition to  CSM interaction if a significant amount of $^{56}$Ni is produced, although CSM interaction by itself can explain SN~2006gy \citep{miller2010sn2006gyir}.

\section{Other models}\label{sec:othermodels}
% added by TM on 14 Aug 2017
Several other ideas have been raised as the power sources of SLSNe. For example, the fallback accretion onto the central compact remnant of the exploding stars can potentially release a sufficient energy to power SLSNe \citep{dexter2013fallback}. It is also suggested that a latent heat released by the phase transition from neutron stars to quark stars is sufficient to power SLSNe \citep[e.g.,][]{ouyed2012quark}.

It is possible that not a single energy source is playing a role in powering SLSNe. Two or more energy sources can provide heat sources at the same time.

We also need to keep in mind the possibility that none of the energy sources currently proposed are correct -- a power source that is completely missed so far may still exist.

%\section{Summary}

\begin{acknowledgements}
This review made use of the Weizmann interactive supernova data repository - \url{http://wiserep.weizmann.ac.il} \citep{yaron2012wiserep} and the Open Supernova Catalog - \url{https://sne.space/} \citep{guillochon2017opensncat}.
TJM is supported by the Grants-in-Aid for Scientific Research of the Japan Society for the Promotion of Science (16H07413, 17H02864) and the Munich Institute for Astro- and Particle Physics (MIAPP) of the DFG cluster of excellence "Origin and Structure of the Universe." The work of ES (interaction models) is supported by the
Russian Scientific Foundation grant 16--12--10519. 
RAC was supported in part by NASA grant NNX12AF90G.
\end{acknowledgements}

%\section{Section title}
%\label{sec:1}
%Text with citations \cite{RefB} and \cite{RefJ}.
%\subsection{Subsection title}
%\label{sec:2}
%as required. Don't forget to give each section
%and subsection a unique label (see Sect.~\ref{sec:1}).
%\paragraph{Paragraph headings} Use paragraph headings as needed.
%\begin{equation}
%\end{equation}
%
%% For one-column wide figures use
%\begin{figure}
%% Use the relevant command to insert your figure file.
%% For example, with the graphicx package use
%  \includegraphics{example.eps}
%\end{figure}
%%
%% For two-column wide figures use
%\begin{figure*}
%% Use the relevant command to insert your figure file.
%% For example, with the graphicx package use
%  \includegraphics[width=0.75\textwidth]{example.eps}
%\end{figure*}
%\begin{table}
%\begin{tabular}{lll}
%\hline\noalign{\smallskip}
%\end{tabular}
%\end{table}

%\begin{thebibliography}{}
%\bibliographystyle{spbasic}
%\bibliographystyle{aa}%{apj}
%\bibliographystyle{plain}
\bibliographystyle{spbasic}
\bibliography{foreword15}
%\end{thebibliography}

%\bibliographystyle{aps-nameyear}      % American Physical Society (APS) style, author-year citations
%\bibliographystyle{aa}
%\bibliography{ferrari}                % name your BibTeX data base
%\nocite{*}
%{\bf References}

%\bibliographystyle{aps-nameyear}
%\bibliography{ferrari}                % name your BibTeX data base
%\nocite{*}

% Format for proceedings
%\bibitem[\protect\citeauthoryear{Oz and Yannakakis}{1983}]{RefB}
%W.V. Oz, M. Yannakakis (eds.),
%in \textit{All ACM Conferences} (Academic Press, Boston, 1983).
%This is a full PROCEEDINGS entry
% Other formats available: INPROCEEDINGS, PHDTHESIS, TECHREPORT,
% UNPUBLISHED, MISC, MASTERSTHESIS, MANUAL, INCOLLECTION, BOOKLET
% etc
%\end{thebibliography}

\end{document}